\title{Topological Voxelization and Graph Generation from 3D Images}
\author{Pirouz Nourian, Shervin Azadi, Samaneh Rezvani}

\documentclass[10pt]{article} 

\usepackage{geometry} 

\usepackage{indentfirst} 
\usepackage{microtype}

\usepackage{times}

\usepackage{fancyhdr,graphicx,amsmath,amssymb, mathtools}
\usepackage{euscript} 

\DeclareFontFamily{OT1}{pzc}{}
\DeclareFontShape{OT1}{pzc}{m}{it}{<-> s * [1.10] pzcmi7t}{}
\DeclareMathAlphabet{\mathpzc}{OT1}{pzc}{m}{it}

\DeclarePairedDelimiter{\nint}\lfloor\rceil 
\usepackage[table]{xcolor}
\usepackage[
	linesnumbered,
	boxed,
	ruled,
	vlined
]{algorithm2e}

\SetCommentSty{mycommfont}

\SetNlSty{mynumfont}{}{}
\SetKwFor{While}{while}{do}{end}
\usepackage{setspace}

\usepackage{enumitem}

\usepackage{multirow,tabularx, booktabs} 

\usepackage[justification=centering, 
font=small, 
labelfont=bf, 
textfont=it]{caption} 

\newcounter{magicrownumbers}

\usepackage{float} 
\usepackage{placeins} 
\usepackage{subcaption} 

\usepackage{hyperref} 
\usepackage[
    backend=biber,
    style=apa
]{biblatex} 

\usepackage[toc,title]{appendix} 

\newcommand{\comment}[1]{} 

\usepackage{lineno}
\nolinenumbers 

\begin{document}
\begin{center}
	{\huge \bfseries Voxel Graph Operators}
	\vspace{0.5cm}
	\\
	{ \large Topological Voxelization, Graph Generation, and Derivation of Discrete Differential Operators from Voxel Complexes}
	\vspace{0.5cm}
	\\
	{\bfseries Pirouz Nourian and Shervin Azadi}
\end{center}

\vspace{0.5cm}

\begin{figure}[!htbp]
	\centering
	\includegraphics[width=\textwidth]{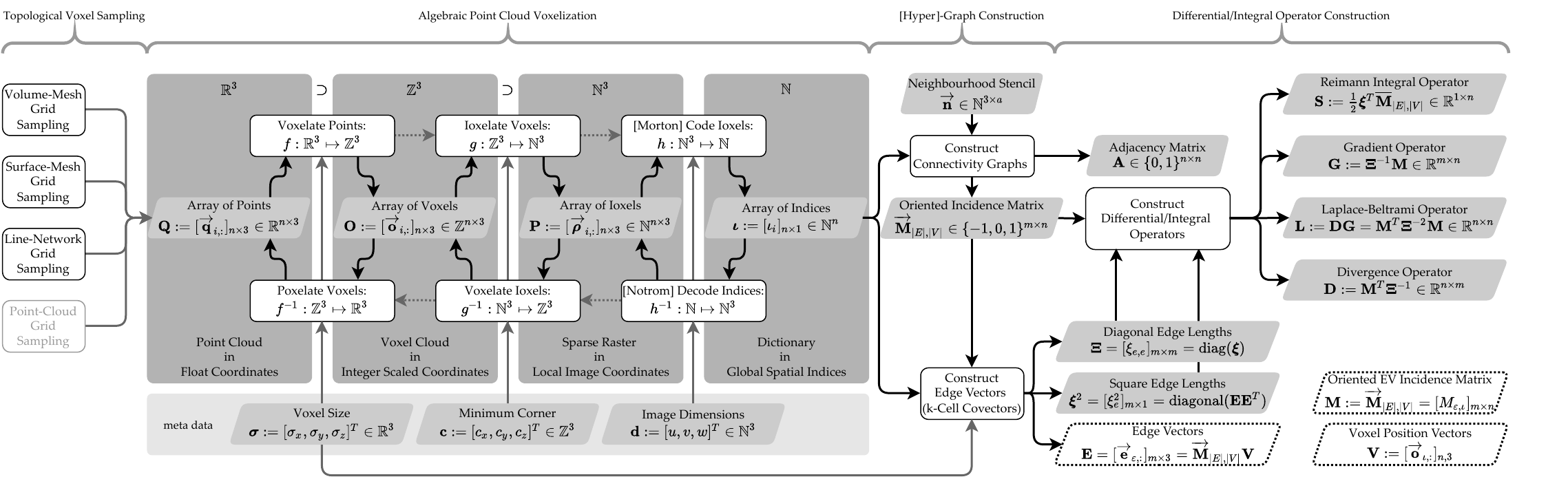}
	\label{fig:workflow}
	\caption*{graphical abstract: a workflow for topological voxelization, graph construction, and differential/integral operator construction
	}
\end{figure}

\begin{abstract}

	In this paper, we present a novel workflow consisting of algebraic algorithms and data structures for fast and topologically accurate conversion of vector data models such as Boundary Representations into voxels (topological voxelization); spatially indexing them; constructing connectivity graphs from voxels; and constructing a coherent set of multivariate differential and integral operators from these graphs. Topological Voxelization is revisited and presented in the paper as a reversible mapping of geometric models from $\mathbb{R}^3$ to $\mathbb{Z}^3$ to $\mathbb{N}^3$ and eventually to an index space created by Morton Codes in $\mathbb{N}$ while ensuring the topological validity of the voxel models; namely their topological thinness and their geometrical consistency.
	In addition, we present algorithms for constructing graphs and hyper-graph connectivity models on voxel data for graph traversal and field interpolations and utilize them algebraic in elegantly discretizing differential and integral operators for geometric, graphical, or spatial analyses and digital simulations.
	The multi-variate differential and integral operators presented in this paper can be used particularly in the formulation of Partial Differential Equations for physics simulations. 


\end{abstract}

\section{Introduction}
Raster data models or images are discrete data models based on a regular grid of small elements: pixels in 2D images and voxels in 3D images.
Raster data models can be advantageous in certain operations and queries in Geospatial Sciences, Spatial Analysis, Scientific Visualization, Medical Imaging, Computer Graphics, and Engineering Optimization.

Central to these applications is working with discrete representations of functions on a network or a manifold object possibly of high genera (many holes), making spatial analysis highly complex.
In medical imaging, voxel data is often the natural source of data generated directly from MRI or CT scans, which is already a convenient representation for spatial analysis.
In scientific simulations and engineering optimization, however, it is sometimes necessary to convert the so-called vector data models of manifold spatial objects in the form of points, curves, surfaces, or solids into voxels for various reasons, namely discrete simulations such as those solved by the Finite-Difference Method, removing noise from data, or in general for regularizing some geometry processing tasks.
Keeping the topological invariants of the input manifold objects in such conversions intact is a challenging task.

For example, converting a point cloud to a raster model can simplify neighbour search and segmentation by reducing confusion and irregularities in point clouds.
The significant advantage of raster representation of geometric objects can be described as unifying geometrical and topological data into one representation.
This means that with a raster model of an object, one can not only decipher geometric shape from the position of the voxels but also discern the topological properties from the neighbourhood of the voxels in question.
In doing so, one can perform almost all geometry processing operations as vectorized (i.e. algebraic) operations with matrices and vectors.
This is highly advantageous for modern computing practices as it increases the potential for parallelization, e.g. in treating geometric data as image data or graph data for Machine Learning or Deep Learning applications.
However, rasterization can potentially corrupt some of the topological information contents that exist in the original vector data models, especially if done without an explicit idea of topological adequacy.
This paper presents a systematic workflow for topological voxelization and voxel-graph construction that can provide confidence, efficiency, and elegance in dealing with geometry processing problems using raster representations.
The outlook of utilizing this workflow is to convert some complicated or error-prone geometry processing problems in computational geometry to integer computing problems in computational topology that do not suffer from floating precision issues common to computational geometry problems.
For example, problems such as Voronoi tesselations or Agent-Based Modelling can be easily converted into topological and combinatorial problems by approximating the input geometric domains as 3D images.

In particular, we consider some essential meta-level use-cases of voxel-representations and voxel graphs, such as solving Partial Differential Equations (PDE) and the typical interpolations in the numerical methods of scientific computing (q.v. \cite{LeVeque2007FiniteDifferenceMethods}).
More specifically, one of the paper's contributions is a comprehensive and algebraic characterization of interpolation, differential, and integral operators on voxel graphs and their utility for solving Partial Differential Equations (PDE) in scientific computing.
These algebraic differential/integral operators are based on some proposed hyper-graph data models to be constructed based on raster data models.
One fundamental use-case of these hyper-graphs (i.e. graphs whose edges can be higher-dimensional cells, a.k.a. topological cell complexes) is to simplify the vectorization or visualization of level-sets in ordinary graphics pipelines for creating iso-surfaces in regular 3D images and 3-manifolds or for creating iso-curves on flat pictures and manifold surfaces by facilitating the use of algorithms such as Marching Cubes and Marching Tetrahedra \cite{Osher2001LevelSetMethods}.
Note that iso-curve and iso-surface creation has remained an active area of research, firstly due to the need for creating alternatives to the patented Marching Cubes algorithm and later for enhancing precision or dealing with irregularities, see, e.g. the Dual Contouring by \cite{ju2002dual}, Surface Nets by \cite{gibson1998constrained}, Manifold Dual Contouring by \cite{schaefer2007manifold}, 2.5D Dual Contouring for building model reconstruction from aerial point-clouds by \cite{zhou20102} and Neural Dual Contouring for the challenge of surface reconstruction from noisy data by \cite{chen2022neural}.
While these use-cases require hyper-graphs, a more straightforward yet less frequent use-case is conducting graph-traversals such as optimal path algorithms, neighbourhood search, sorting, random-walk/diffusion processes \cite{nourian2016configraphics}.

In short, this paper presents a novel computational workflow consisting of compatible data structures and algorithms for efficiently converting piece-wise linear geometric data into topologically adequate voxel data and constructing hyper-graphs from voxels.
The proposed pipeline is an algebraic, vectorized, efficient, and explicitly topological modelling process that can be easily generalized to higher-dimensional data analytics procedures for use cases in Topological Data Analysis (TDA) \cite{zomorodian2012topological}.
The efficiency of the algorithms is claimed in terms of minimal space [memory] and [computing] time complexity.
Additionally, the adequacy or efficacy of processes is claimed in terms of topological correctness and sufficiency, i.e. preserving such topological properties as continuity, closeness, and thinness.

In the following sections, we first highlight notable works in the same direction to illustrate this work's background and then elaborate on our framework by describing the relationship between the mathematical objects used in this pipeline.
The central part of this paper is the methodology section, where we explain all the steps in the proposed pipeline: mesh sampling, point cloud rasterization, construction of proximity graphs, and construction of discrete differential/integral operators.
Lastly, we demonstrate the advantages of this novel approach by showing the results of a $L_p$ Voronoi tessellation of a 3D manifold based on a diffusion distance obtained by random-walk simulations on a graph obtained from synthetic voxel data.
Additionally, we present the open-source computational implementation of a significant part of the proposed pipeline in the topoGenesis python package \cite{azadi2020topogenesis}.
The notational conventions of the paper are explained in Table \ref{tab:NomenClature} in Appendix A.

\section{Background}
In this section, we cover the general context of rasterization and voxelization methods;
we look into state of the art in voxelization algorithms that explicitly aim to maintain the topological properties of the original model;
and finally, we elaborate on the two major application domains that this pipeline can contribute to:
(1) ensuring geographical-topological consistency in geo-spatial analysis and geospatial simulations.
(2) alleviating the notorious challenges of mesh generation and mesh dependency in numerical methods, e.g. solving Partial Differential Equations in Finite Element Method (FEM) or Spectral Methods.

\subsection{Voxelization}
Since the advent of raster displays, rasterization became a fundamental process in modern computers for visualizing objects on digital screens (q.v. \cite{Uno1979AGeneralPurpose, Amanatides1987FVT}).
The regularity of raster representations makes them more suitable for mathematical modelling and analyses.
Therefore, rasterization of so-called \emph{vector geometry} inputs is a common step in more complex computational procedures.
Principally, voxelization is the volumetric generalization of the rasterization process that aims to convert vector representation of geometry data into volumetric pixels, an operation typically referred to as Scan Conversion.

Various voxelization procedures directly inspired by the Scan Conversion have been presented in the computer science literature, the oldest of which date back to the 1980s.
Most early works in this area are based on the idea of making a Digital Differential Analyzer (DDA), i.e. an algorithm that converts the slope of lines from a float to an integer quotient describing the rhythmic number of steps in $\rm{X, Y, Z}$ directions to voxelate them efficiently.
The most prominent example of which is the Fast Voxel Traversal method by \cite{Amanatides1987FVT} in that it was a visionary extension of the idea of the DDA to 3D voxelization of lines.
Notably, another method for Scan Conversion-based DDA was proposed in the same year for voxelization of triangles by \cite{kaufman_kaufman_1987}.
This algorithm is particularly important as triangle meshes are the most common data models for representing digital surfaces and the boundary representation of solids or 3-manifolds.
Adding the introduction of the Marching Cubes by \citeauthor{lorensen1987marching} also in 1987 marks the year 1987 as the de facto birthdate of voxelization in Computer Graphics, i.e. 35 years ago.

The DDA approach to voxelization is followed by \cite{kaufman_efficient_1988, cohen-or_fundamentals_1995} mainly focusing on direct voxelization of geometric primitives using their parametric definitions (more recently in \cite{hutchison_voxelization_2006}).
In image-based approaches to voxelization of solids (3-manifolds), the bounding box of the geometric model is first sliced into 2-dimensional images.
As proposed by \citeauthor{yuen-shanleungConservativeSamplingSolids2013}, the images then are filled by shooting rays perpendicular to the slicing direction (\cite{yuen-shanleungConservativeSamplingSolids2013}).
To increase the efficiency of the costly intersection computations, \cite{youngGPUacceleratedGenerationRendering2018} generalized the image-based approach to perform a multi-level voxelization.
Similarly, the Octree structure can be used to minimize the computational cost.
Namely, in \cite{crassinOctreeBasedSparseVoxelization2012} the authors project each triangle to the dominant side of it instead of shooting rays to form a sparse voxel model based on Octree.
With the same aim, \cite{kampeHighResolutionSparse2013} proposed a new data structure for voxels based on a directed acyclic graph (DAG) instead of Sparse Voxel Octree.
Despite their efficiency and generality, none of these methods can guarantee the preservation of the topological properties of the geometric model during the conversion.
See a recent literature review on voxelization algorithms and data structures by \cite{aleksandrov2021voxelisation}.

\subsection{Topological Properties}
Explicit attention to topological accuracy can be traced back to \cite{huangAccurateMethodVoxelizing1998} followed by a seminal book on digital geometry with various chapters dedicated to topological connectivity in voxel datasets by \cite{klette2004digital}, a concise mathematical introduction to the matter by \cite{laine_topological_2013}, and an algorithmic definition by \cite{nourian_voxelization_2016}.
\cite{laine_topological_2013} presents the mathematical ideas of topological correctness of voxelated shapes and presents the mathematical essence of topological voxelization but does not explicitly present algorithms for voxelization of input geometry.
\cite{nourian_voxelization_2016} present algorithms for 0D, 1D, 2D, and 3D inputs, respectively, for voxelating point clouds, curves, surfaces, and solids.
Although these algorithms are implemented and tested for processing geospatial data, they need to be improved in terms of time complexity and output data structures.
The complexity of the algorithms presented in that paper is in the order of $\rm{O(UVW)}$ where $\rm{U, V, W}$ respectively represent the number of voxels in $\rm{X, Y, Z}$  directions.
For comparison, here we present an algorithm with the complexity of $\rm{O(UV+VW+WU)}$, which consists of three procedures arguably comparable in efficiency to the Scan-Conversions by \cite{cohen-or_fundamentals_1995} and GPU computing approaches (e.g. the one by \cite{youngGPUacceleratedGenerationRendering2018}) for pixelating inputs while offering explicit control on connectivity levels.

\subsection{Geospatial Consistency}
Many of the physical behaviours and correlated features of the modelled phenomena heavily depend on the notion of closeness that is naturally represented by the topological properties of data and models.
Loss of global topological properties such as connectedness \& closure during the geospatial data modelling process can be detrimental to the usability of the final results of the spatial analysis and simulation procedures.
Preservation of the global topological properties ultimately requires consistency at the level of the local topology of the digital spatial objects that are defined based on the notion of adjacency between the cells of the background space.
Additionally, maintaining geographic consistency in the modelling process requires the preservation of topological relations of objects as well as the existence of a reversible geospatial indexing schema to ensure that tiled or dissected models can be collated consistently together to make models of larger areas.
When integrating different geographical data sources, it is essential to measure the similarity of objects, e.g. a road represented by a polyline in one map and the same road represented by a polygonal region in another, by comparing their topological properties at a certain level of detail \cite{Belussi2005TowardsTopologicalConsistency}.

The issue of topological consistency is not unique to raster representation.
In fact, it can be more challenging to preserve topological consistency in workflows requiring multi-scale vector data that needs to be generalized (lowering the geometric level of detail in the geospatial analysis is referred to as generalization), as explained by \cite{vander_Poorten_2002TopologicallyConsistentMap}.
To provide for a consistent query structure across vector and raster data models, \cite{Egenhofer1993TopologicalRelationsBetween} investigated the similarity of topological properties in vector and raster data; proposing that the topological properties in $\mathbb{R}^2$ are a subset of topological properties in $\mathbb{Z}^2$.
Correspondingly, there have been various approaches to the unification of the notion of topological properties between raster and vector data, such as the mathematical work by \cite{Winter2000} and the work on data models by \cite{Voudouris2010TowardsUnifyingFormalisation}.
While the term geospatial usually has the connotation of referring to outdoor large-scale 2D domains, a relatively new line of research on 3D indoor models of large and complex buildings shows the importance of geospatial consistency in small-scale 3D spatial analyses, see, e.g. this work on indoor navigation by \cite{gorte2019navigation}.


\section{Proposed Framework}
The central point of the proposed framework is the reintroduction of graphs and [regular] cell complexes as locally-linear and globally non-linear discrete models of manifolds.
This is not only for discrete geometry processing methods (see this seminal book by \cite{klette2004digital}) to benefit from image processing techniques (see this presentation \cite{chen2004discrete}) but also for Topological Data Analysis (see a topology oriented introduction by \cite{zomorodian2012topological}, and an applied statistics-oriented introduction by \cite{wasserman2016topological}), Manifold Learning (see an introduction by \cite{izenman2012introduction}), and algorithm design for algebraic Simulations (in particular for solving Partial Differential Equations (PDE), e.g. using graph theoretical formulations of the Finite Element Method similar to those proposed by \cite{christensen1988network}.)

In this regard, we reflect on the general utility of graphs as \emph{interpolation objects} that provide locally linear interpolation spaces for manifold spaces that are globally non-linear and thus irreducible to linear spaces.
The utility of graphs and simplicial complexes as discrete interpolation objects (or approximate manifolds) must be evident due to the guaranteed linearity of k-simplexes; however, 2D quadrilateral cells or 3D hexahedral cells might be non-planar in general and thus require more sophisticated interpolations
On the other hand, k-cubes (interpolation objects in voxel graphs) are rendered planar due to the regularity of their grids, thus admitting multilinear interpolations on par with barycentric interpolations in simplexes (see Table\ref{tab:topologicalPrimitivesInterpolation} for a summary.)

\begin{table}
	\centering
	\footnotesize
	\caption{Piece-wise Linear Data Models for approximating non-manifold and manifold spatial domains in which fields can be sampled, interpolated, and analysed, note that non-manifold spatial domains can be generally represented by point-clouds or voxel-clouds as simplicial/multi-linear complexes}
    \begin{tabularx}{\textwidth}{
    >{\raggedleft\arraybackslash\hsize=1.32\hsize}X
    >{\raggedright\arraybackslash\hsize=0.62\hsize}X|
    >{\raggedright\arraybackslash\hsize=1.02\hsize}X
    >{\raggedright\arraybackslash\hsize=1.02\hsize}X
    >{\raggedright\arraybackslash\hsize=1.02\hsize}X
} 
    \toprule
    \multirow{2}*{Representation Type} 
    & Field Object
    & \multicolumn{3}{>{\raggedright\hsize=1.02\dimexpr3\hsize\tabcolsep+\arrayrulewidth\relax}X}{
        combinatorial topological objects for constructing graphs and/or cell complexes 
        
    } 
    \\ 
    \cline{2-5}
    & Field Samples
    & 1D-Manifold \linebreak 1D Interpolation 
    & 2D-Manifold \linebreak 2D Interpolation
    & 3D-Manifold \linebreak 3D Interpolation
    \\ 
    \toprule
    K-Cube Objects
    & Voxel
    & Segment 
    & Square \linebreak (or 2 triangles) 
    & Cube \linebreak (or 6 tetrahedrons)
    \\ 
    Regular Complexes, \linebreak Multi-Linear Interpolations
    & $\rightarrow$  
    & Linear 
    & Bilinear 
    & Trilinear
    \\ 
    Regular Networks \comment{or Cartesian Complexes, all \emph{embedded in $\mathbb{Z}^3$}, Lattices of k-cubes}
    & Voxel Cloud 
    & Raster Curve
    & Raster Surface
    & Raster Volume
    \\ 
    K-cube Level-Sets (Contours) 
    & 
    & Marching Lines: \linebreak Contour Points
    & Marching Squares: \linebreak Contour Curves
    & Marching Cubes: \linebreak Contour Surfaces
    \\ 
    \toprule
    K-Simplex Objects
    & Cell
    & Line-Edge
    & Tri-Face 
    & Tet-Cell
    \\ 
    Irregular Simplices, \linebreak Barycentric Interpolations
    & $\rightarrow$
    & 1-Simplex 
    & 2-Simplex 
    & 3-Simplex
    \\ 
    Irregular Networks \comment{or Simplicial Complexes all \emph{embedded in $\mathbb{R}^3$}, Boundary Representations made up of k-simplexes}
    & Point Cloud 
    & Line Network 
    & Surface Mesh 
    & Volumetric Mesh 
    \\ 
    K-Simplex Level-Sets (Contours) 
    &
    & Marching Lines: \linebreak Contour Points
    & Marching Triangles: \linebreak Contour Curves
    & Marching Tetrahedra: \linebreak Contour Surfaces
    \\ 
    \toprule 
    Intersection Targets 
    & Hyper-Plane \linebreak 3D Intervals
    & Plane \linebreak 2D Intervals
    & Line \linebreak 1D Intervals
    & Point \linebreak 0D Intervals
    \\ 
    \toprule 
\end{tabularx}
	\label{tab:topologicalPrimitivesInterpolation}
\end{table}

To recapitulate, the local linearity of graphs and hyper-graphs or topological cell complexes in their edges or k-cells (k-dimensional finite elements such as surface elements and volume elements) is a major advantage for a wide range of algorithms in that it allows for breaking down a complex and non-linear space into bounded linear subspaces whose data models can be represented simply as tuples of integers, literally.
While we have chosen to emblematically reintroduce graphs and hyper-graphs (a.k.a. topological cell complexes) as to their utility for interpolations, in fact, their utmost utility is for elegantly defining sparse differential and integral operators, specifically discrete differential operators such as the gradient, divergence, and Laplace-Beltrami operator or Riemannian integral operators.

What all the applications listed above have in common is that one often needs to work with a finite number of samples of a scalar-valued or vector-valued function (also known as a scalar field or a vector field.)
These samples are typically extracted from a non-trivially shaped spatial region, where there is an assumption that this region (known as the [input] domain of the function) is a d-dimensional manifold (typically a 2-manifold or 3-manifold).
A d-manifold is a globally [non-linear] region in a Euclidean space that is locally homeomorphic or similar to a Euclidean space of dimension $d$ or $\mathbb{R}^d$, thus locally linear and smooth enough such that its tangent spaces at small-enough neighbourhoods can be approximated by linear spaces such as simplexes (1D line-segments, 2D triangles, or 3D tetrahedrons) or [straight] k-cubes (1D line-segments, 2D squares, 3D cubes).
This means that an arbitrary point in the manifold cannot be expected to be constructible from an interpolation of a finite number of points from the manifold in general.
However, the manifoldness means that in the small neighbourhoods of the points in the manifold region, such linear interpolations would cover the local neighbourhood space, i.e. every point in the local neighbourhood can be addressed and obtained with interpolation parameters in a fashion similar to coordinates in a Euclidean space.

The so-called \emph{piece-wise} linear geometric models are indispensable for many engineering and scientific computing applications concerned with PDE.
This is common, inter alia, in Finite Element Analysis (FEA), where some quantities pertaining to a finite set of vertices are computed as solutions to a PDE, which can then be interpolated in the finite surface or volume elements between the vertices for obtaining a smooth picture of the field in question.
This makes it possible to compute the results of the equations on a finite set of points while having the opportunity to interpolate and obtain smooth results as required per application.
Note that the interpolation objects defined here on digital images (pixel grids or voxel grids) are dual to the k-cells of the discrete manifold objects in question, unlike simplicial complexes where the k-cells are the same as interpolation objects (i.e. triangles of the marching triangles will be the same as faces and so forth).
To understand this, loosely speaking, consider the Poincaré duality of the combinatorial cube objects in the Marching Cubes algorithm by \cite{lorensen1987marching} to the voxels in the raster (digital 3D image) or the duality of the combinatorial square objects to the pixels in the Marching Squares algorithm (see a related introduction to the concept of duality by \cite{lee2005combinatorial}).

In TDA, interpolation (and extrapolation) can be considered the ultimate aim of the analytic procedure with the objective to predict the properties of data points between or around those sampled before.
To interpolate scalar/vector fields between the spatial data points, there must exist some topological objects of higher dimensions.
These higher-dimensional topological objects are best understood in the language of Algebraic Topology (a.k.a. Combinatorial Topology see the definition in \cite{zomorodian2012topological} or \cite{hatcher2005algebraic}).

In the following subsections, we describe the importance of topological properties of manifolds, the discrete way to model or approximate them with combinatorial objects, what these objects are with regards to the bigger picture of the global topological properties of the manifold space, as well as the topological similarity of these manifolds to such things as disks, washers, balls, tori, and alike, as described in the Euler-Poincaré characteristic of their discrete representations.
Finally, the last subsections will present the basis of the proposed methods for graph construction and the derivation of the differential/integral operators on voxel graphs.


\subsection{Point Set Topology \& Algebraic Topology}
In the two broad categories of use cases introduced above as TDA (particularly spatial data analysis) and spatial Simulations (i.e. first principles formulated as PDEs), the topology of the object under study is key to the way it conducts flows; be it the flow of fluids, heat, forces, pedestrians, or electrical current.
Such flows are core concepts in defining the system's state in simulations, or the geodesic distances through the manifold that shapes the basis of \emph{spatial analysis} and set it apart from other data analyses.
Evidently, the way a manifold conducts or resists (admits or impedes) such flows depends on its topology or network structure as to which essential analogies between graphs and electrical circuits or hydraulic networks can be made to utilize the fundamental theorems of Kirchhoff for computing flows in circuits. 

In practice, the prerequisite of flow analysis or flow simulation is constructing a topological picture of theoretically continuous manifold spaces using discrete data points for making networks (graphs) or discrete manifolds (cell complexes).
A common challenge is the correspondence of the global topological properties of the discrete manifold model with the supposedly continuous domain it models.
A key concept that connects the seemingly disparate worlds of continuous manifolds and discrete manifolds is the notion of a locus or a point-set as defined in terms of a crisp membership function or a predicate defining the explicit (i.e. based on analytical equations for the coordinates) or implicit (i.e. a level-set description of a function only dealing with a predicate pertaining to the scalar or vector attributes of points) characteristics of points belonging to the point-set.
Two definitions of topology are particularly relevant to the discussion here: one concerning the topological properties of the global picture of the objects in terms of such things as their genera \comment{(plural form of genus or handle, e.g. the two handles of a double torus)}, inner shells, and alike, as well as another one concerning the topological constructs building a discrete picture of the local neighbourhoods in the computational representation of the objects.
The first sense is, in fact, General Topology or Point-Set Topology, which regards objects as point-sets or loci. 
Of course, the point-set topological properties of single objects affect the spatial relations between such spatial objects, e.g. see a paper by \cite{egenhofer1991point} on the subject of spatial relations between spatial regions in 2D, and a comprehensive introduction to the 3D cases by \cite{zlatanova2017topological}.
These properties, however, are known to be scale-dependent, as evident in the idea of Persistent Homology (the core of TDA in Computational Topology, see an explanation by \cite{edelsbrunner2022computational}), which can be described as the study of the topological \emph{big picture} of a data set, in a manner of speaking.
Mathematically, the topological big picture of a manifold spatial region is summarised and coded into its Euler-Poincaré characteristic, which in the case of 2-manifolds without borders or cavities is defined as:

\begin{equation}
	\chi(\EuScript{M})=V-E+F=2-2g.
\end{equation}

Leonhard Euler originally proposed the left-hand side (LHS) of this equation for describing the topology of polyhedrons, referring for the first time to the combinatorial topological constructs (vertices, edges, and faces) constituting the digital model of the object (an algebraic topology picture).
Henri Poincaré added the right-hand side (RHS) to refer to the global topological properties of the object, such as the number of genera/tunnels (denoted as $g$), cavities or shells (not present in this simple formula), thus describing a picture of the entire point set topology as a locus.
Naturally, the LHS is easier to check and obtain for a computer, while the RHS is supposedly easier to compute visually for a human.
Ascertaining the equality of the two, i.e. the supposed value of the RHS with the actual value of the LHS, is a matter of topological validation of data.
See an extended version of the equation and ways to compute the challenging RHS for validating the topology of voxelated manifolds as proposed by \cite{sanchez2013euler}, a thorough introduction to the ideas of topological validity and sufficiency in \cite{weilerEdgeBasedDataStructures1985}, \cite{weiler1986topological}, and a comprehensive introduction to the generalization of the idea to non-manifold objects by \cite{masuda1993topological}.

The prerequisite of understanding such a topological big-picture in a digital setting is that one needs to have made a simplicial complex or a topological cell complex of the dataset that contains algebraic or combinatorial topological objects known as k-cells (see \ref{tab:topologicalPrimitivesInterpolation}\comment{a generalization of the topological terms 0D vertex, 1D edge or curve element, 2D face or surface element, and 3D cell or volume element}) defining the edges or hyper-edges as interpolation objects between data points.
In this sense, simplicial complexes such as the Vietoris-Rips Complexes or Alpha Complexes generalize the idea of graphs to hyper-graphs.

In summary, it suffices to say that without an implicit notion of \emph{natural topology} amongst some data points, they together represent nothing more than a shape less \emph{powder}.
It is precisely this notion of topology that allows one to conceive of the existence of points in between sampled points.
However natural or trivial this notion of topology may seem, this intuition about the existence or meaningfulness of points in between points can lead to mistakes in dealing with non-trivially shaped spatial regions or manifolds, which are locally Euclidean thus leading our intuition in this direction but globally non-Euclidean.
An archetypical example of such common mistakes is the assumption that two points whose coordinates are close in the Euclidean/Cartesian sense are actually close in space while they might be on the two sides of a river and only connected through a non-trival manifold of roads of high genera (with many holes, so to speak) thus actually far apart from the sense of geodesic flows.
TIn the following sections we shall see how the proposed workflow can simplify the explicit construction of models of topology for unambiguously describing the connexions between voxels; explicit models defined as incidence and adjacency matrices representing bipartite or unipartite graphs explicitly encoding topological links amongst the k-dimensional cells of topological cell complexes made up of voxels (see the Section \ref{sec:GraphConstruction} for algorithmic details,  Table\ref{tab:CombinatorialGraphs} for a summary, and Section\ref{sec:DiscreteVectorCalculus}).


\subsection{Topological Complexes As The Interpolation Space}

The so-called k-cells are combinatorial constructs defined in algebraic topology (q.v. \cite{hatcher2005algebraic}), which are introduced under the umbrella of computational topology, see a short introduction by \cite{zomorodian2009computational}, and two books by \cite{zomorodian2005topology} \& \cite{edelsbrunner2022computational}.
The cells introduced in this paper are voxel links as line segments, squares of the marching squared the cubes of marching cubes (see Table \ref{tab:topologicalPrimitivesInterpolation}).

Combinatorial topological cell complexes can be made by \emph{glueing together} linear cells, hence the name piece-wise linear representations [of regions in space].
This framework suggests utilizing combinatorial graphs constructed from the commonalities (adjacency-relations between same-dimensional cells or incidence relations between different-dimensional cells) as a convenient replacement for the so-called \emph{natural topology} of the Euclidean space (open balls or open intervals).
While the construction of a navigable topological object (a graph or hyper-graph) in the Euclidean space is generally hard or ambiguous, as elaborated in the next section, in the completely discretized world of voxels, such graphs or hyper-graphs can be easily and unambiguously constructed algebraically from incidence matrices as proposed in Table \ref{tab:CombinatorialGraphs}.

Graphs are the most common and versatile topological models of spatial regions consisting of only 0D-Vertices and 1D-Edges (geometric graphs are often denoted as an ordered pair of vertices and edges dubbed $G=(V, E), \; E\subset V\times V$).
However, there do also exist higher-dimensional topological models of spatial regions with surface elements or volume elements such as simplicial complexes (triangular or tetrahedral meshes, generally denoted as ordered pairs of vertices, edges, faces, dubbed as $M=(V, E, F), E\subset V\times V, F\subset V\times V\times V$).
The necessity of higher-dimensional k-cells in our proposed framework is, respectively, for interpolation or iso-curve construction on 2D slice pictures of 3D images and iso-surface construction out of 3D scalar fields.
A thorough introduction to such combinatorial objects for meshes can be found in \cite{weilerEdgeBasedDataStructures1985}.

\subsection{Graph Construction}
Constructing proximity graphs or topological cell complexes for approximating manifolds from point samples is generally a difficult task specially on point clouds.
The two famous methods of constructing the so-called epsilon-graphs and k-nearest neighbour graphs are only two examples of a larger family of neighbourhood definitions for constructing proximity graphs (for general purpose TDA as epsilon ball graphs, KNN graphs, Gabriel Graphs, Relative Neighbourhood Graphs beta-skeletons or for specific applications like modelling ad-hoc telecommunication networks with unit disk graphs).
The variety of the methods should already convey the difficulties of obtaining a persistent topological picture from such unstructured data.
On the contrary, as shall be presented in this paper, constructing graphs on voxel data is very elegant and straight forward as compared to point clouds.
In addition to this the Cartesian regularity of voxel graphs makes it easy to decipher geometric information from topological information contents.
This makes voxel graphs particularly appealing for constructing differential/integral operators.

Based on a given voxel cloud, various neighbourhoods can be defined.
In this paper, we propose to use a computational object called \emph{stencil} that represents a graph that can function as a kernel on the voxel cloud.
Like the kernels in the image processing, the stencil describes a set of conditions based on relative indices, which can be checked by moving the kernel on the image, similar to discrete convolution.
However, in contrast to the discrete convolution where the output is a pixel, the stencil also describes a set of edges or hyper-edges that will be constructed if the conditions were true.
Naturally, the edges refer to the relative indices.
Therefore, the graph representing the local topology of the model can be constructed.
Stencils are easily generalizable to higher dimension cells as they are essentially graphs.
Furthermore, stencils can be customized to represent different local topologies to fit the particular application case.

\begin{table}[H]
	\centering
	\caption{Combinatorial Graphs between topological primitives up to dimension 3: the diagonal entries represent adjacency relations among complexes of the same dimension, whereas the off-diagonal entries represent incidence relations among complexes of different dimensions.}
    \begin{tabularx}{\textwidth}{|
    >{\small\raggedleft\arraybackslash\hsize=1.32\hsize}X|
    >{\small\centering\arraybackslash\hsize=0.92\hsize}X|
    >{\small\centering\arraybackslash\hsize=0.92\hsize}X|
    >{\small\centering\arraybackslash\hsize=0.92\hsize}X|
    >{\small\centering\arraybackslash\hsize=0.92\hsize}X|
    } 
    \hline
    Combinatorial Graphs 
    & 0D-Vertexes 
    & 1D-Edges 
    & 2D-Faces 
    & 3D-Cells
    \\ 
    \hline
    0D-Vertexes
    & $\mathbf{A}_{VV}$ \cellcolor{gray}  
    & $\overrightarrow{\mathbf{M}}_{VE}$ \cellcolor{lightgray}
    & $\mathbf{M}_{VF}$ 
    & $\mathbf{M}_{VC}$ 
    \\ 
    \hline 
    1D-Edges
    & $\overrightarrow{\mathbf{M}}_{EV}$ \cellcolor{lightgray}
    & $\mathbf{A}_{EE}$ \cellcolor{gray}
    & $\overrightarrow{\mathbf{M}}_{EF}$ \cellcolor{lightgray}
    & $\overrightarrow{\mathbf{M}}_{EC}$ \cellcolor{lightgray}
    \\ 
    \hline 
    2D-Faces 
    & $\mathbf{M}_{FV}$  
    & $\overrightarrow{\mathbf{M}}_{FE}$ \cellcolor{lightgray}
    & $\mathbf{A}_{FF}$ \cellcolor{gray}
    & $\overrightarrow{\mathbf{M}}_{FC}$ \cellcolor{lightgray}
    \\ 
    \hline 
    3D-Cells
    & $\mathbf{M}_{CV}$  
    & $\overrightarrow{\mathbf{M}}_{CE}$ \cellcolor{lightgray}
    & $\overrightarrow{\mathbf{M}}_{CF}$ \cellcolor{lightgray}
    & $\mathbf{A}_{CC}$ \cellcolor{gray}
    \\ 
    \hline
\end{tabularx}
	\label{tab:CombinatorialGraphs}
\end{table}

\subsection{Discrete Vector Calculus Operators for Voxel Graphs}

Here we briefly explain the motivation and the basis for the derivation of the proposed differential/integral operators for voxel graphs.
A particular operator of interest is the versatile Laplacian Operator (a.k.a. the Laplace-Beltrami Operator), known in discrete settings firstly as the Combinatorial Laplacian for graphs that have been extensively discussed in the literature of Spectral Graph Theory \cite{chung1997spectral,spielman2007spectral,nourian2016configraphics}.
Another important and versatile variant of the Laplace-Beltrami operator is the Mesh Laplacian, defined for Computer Graphics applications (mostly \emph{mesh processing}, see \cite{botsch2010polygon}) discrete simplicial (triangular) surfaces \cite{sorkine2005laplacian}.

Spectral Graph Theory (SGT) is mostly focused on the connection between topological or metric properties of graphs and Markov Chains with the spectrum (eigen-pair) of their Laplacian matrices, which have achieved remarkable results in application areas such as Random Walk Simulations, Web Indexing for large-scale search applications, Clustering, Signal Processing, and Machine Learning on graphs.
Spectral Mesh Processing (SMP) focuses on relating common operations on meshes such as smoothing, simplification, segmentation, interpolation, diffusion and so on to the spectrum of the typically cotangent Laplacian defined for discrete triangular surfaces \cite{zhang2010spectral, levy2010spectral}.
Nevertheless, it is rather uncommon in SGT or SMP literature to derive the Laplace operator directly and explicitly from the combination of its constituents, i.e. the Gradient operator and the Divergence operator.
Here, due to our attention to the use of all of these operators in discretizing PDE and solving them, we pay a particular attention to the physical dimension and the physical interpretation of the operators constituting the Laplacian operator.
Additionally, we also define a Riemannian integral operator for discrete functions sampled on voxelated domains.

\section{Methodology}

\begin{figure}[H]
	\centering
	\includegraphics[width=\textwidth]{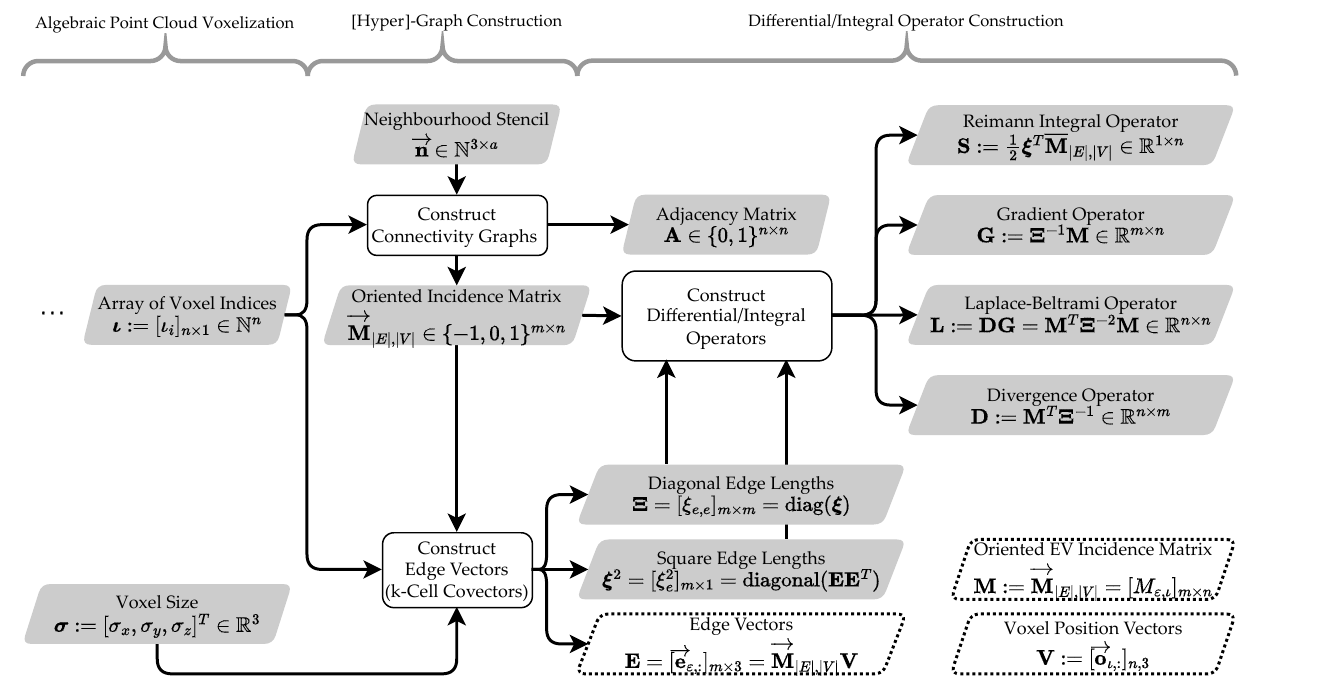}
	\caption{the proposed workflow for Graph Construction \& derivation of Discrete Differential/Integral operators for voxel graphs}
	\label{fig:graphconstruction_operators}
\end{figure}

As it has been outlined earlier, the proposed workflow in this paper consists of four main steps which will be described in the following subsections: Mesh Sampling, Topological Voxelization, Graph Construction, and Derivation of Discrete Differential/Integral Operator.
The Mesh Sampling step is an extension of the topological voxelization idea proposed by \cite{laine_topological_2013} and the algorithms of \cite{nourian_voxelization_2016} which aims to take topologically adequate samples based on the idea of Poincaré Duality between $k$-dimensional objects and $(n-k)$-dimensional objects embedded in an n-dimensional Euclidean space ($\mathbb{R}^n$) from input data models; i.e. extract 0D (point cloud) samples from 3D (volumetric mesh), 2D (surface mesh), and 1D (line networks) manifolds embedded in $\mathbb{R}^3$ ion such way that the corresponding topological properties can be derived from the samples (see Figure \ref{fig:sampling_voxelization}).
This step includes three separate algorithms for each data type: Algorithm \ref{alg:LineNetworkSampling} for line-networks, Algorithm \ref{alg:MeshSurfaceSampling} for mesh surfaces, and Algorithm \ref{alg:VolumetricMeshSampling} for volumetric mesh.
The Topological Voxelization step voxelates, shifts, and encodes the sampled point clouds into voxel clouds that are represented by globally unique spatial Morton indices (see Figure \ref{fig:sampling_voxelization} and Algorithm \ref{alg:PointCloudVoxelization}).
The Graph Construction step derives the topological properties of the original mesh based on a given stencil and encapsulates them in a hyper-graph that can function as the interpolation space (see Figure \ref{fig:graphconstruction_operators} and Algorithm \ref{alg:GraphConstruction}).
The last step derives the discrete integration, differentiation, and interpolation operators on the hyper-graph domain (see Figure \ref{fig:graphconstruction_operators}).



\subsection{Mesh Sampling}
The Mesh Sampling algorithms aim to sample an irregular point cloud from 1D, 2D, or 3D manifolds represented by meshes embedded in $\mathbb{R}^3$.
Thus, in this section, we propose three algorithms to sample a point cloud from Line Networks (Algorithm \ref{alg:LineNetworkSampling}), Mesh Surfaces (Algorithm \ref{alg:MeshSurfaceSampling}), and Volumetric Meshes (Algorithm \ref{alg:VolumetricMeshSampling}) in a manner that retrieving the topological properties of the original mesh is relatively easy in the graph construction step (see Section \ref{sec:GraphConstruction}).
These algorithms can be formalized as a mapping from a manifold domain in $\mathbb{R}^3$ to a point cloud in $\mathbb{R}^3$.
In each of the algorithms, we construct a set of intersection objects with the $(D - d)$ dimensions, where $d$ is the highest number of dimensions in mesh elements and $D = 3$ as the embedding space is $\mathbb{R}^3$.
Later in the process, we will voxelize the point cloud to achieve a voxel cloud (in $\mathbb{Z}^3$) and reconstruct the topological properties of the original mesh from that voxel cloud.
To be able to reconstruct the topological structure of mesh, we need to arrange the intersection objects according to the most basic connectivity type in the stencil used in the graph construction method (see Appendix \ref{tab:ConnectivityType}).

\subsubsection{Line Network Sampling}

The first algorithm describes a Line Network Sampling using planes as intersection objects.
We extend the Ray-Triangle intersection of \citeauthor{Moller_1997FastMinimumStorage} to a line-plane intersection algorithm that reduces the computational cost by exploiting the regularity of the voxel-grid \cite{Moller_1997FastMinimumStorage}.
As \citeauthor{nourian_voxelization_2016} has showed, different intersection objects can capture different kinds of connectivity within the original mesh \cite{nourian_voxelization_2016}.
We need to construct the intersection objects in accordance with the simplest form of connectivity type (see Table \ref{tab:ConnectivityType}) in the stencil that the Graph Construction (Algorithm \ref{alg:GraphConstruction}) uses further down the pipeline.
In the algorithm below, we utilize the \emph{Conservative voxelization}.
Thus the intersection planes are aligned with voxel boundaries; in an example grid of $m$ by $n$ by $o$ voxels, respectively aligned with the $X$, $Y$, and $Z$ axes.
Consequently, we need $m+1$,  $n+1$, and $o+1$ such planes to sample the input line network within a bounding grid.

To extend the algorithm of \citeauthor{Moller_1997FastMinimumStorage}, we need to fnd the intersection of two loci: the line locus ($\mathbf{p}'=\mathbf{p}+r\mathbf{d}, r\in[0,1], d:=\mathbf{v}_1-\mathbf{v}_0$) and plane locus ($\mathbf{p}'=\mathbf{c}+s\mathbf{u}+t\mathbf{v}, s,t\in[0,1]$); thus $\mathbf{p}-\mathbf{c}=-r\mathbf{d}+s\mathbf{u}+t\mathbf{v}$.
Note that here we are looking at a bounded frame made up of two principal vectors of the exact length of the bounded frame in the $u$ and $v$ directions.
Therefore, intersection parameters ($r,s,t$) out of these ranges (i.e. $[0,1]$) will not be acceptable.
We can rewrite our equality algebraically:

\begin{equation}
	[-\mathbf{d},\mathbf{u},\mathbf{v}]_{3\times 3} \begin{bmatrix} r\\ s\\ t\end{bmatrix}=[\mathbf{p}-\mathbf{c}]_{3\times 1}
\end{equation}

This is a system of linear equations in the form of $\mathbf{A}\mathbf{x}=\mathbf{b}$, that we are going to solve analytically by forming the inverse matrix $\mathbf{A}^{-1}$.
With this approach, we can speed up the computation not only because of the closed-form nature of the solver but also because the coefficients of the equation $\mathbf{A}:=[-\mathbf{d}|\mathbf{u}|\mathbf{v}]_{3\times 3}$ (and thus $\mathbf{A}^{-1}$) will be the same for each batch of planes perpendicular to the axes X, Y, and Z of the voxelization domain.

\begin{equation} \label{eq:det}
	\begin{bmatrix} r\\ s\\ t\end{bmatrix}=\mathbf{A}^{-1}[\mathbf{p}-\mathbf{c}]_{3\times 1}=\mathbf{A}^{-1}\mathbf{b}=\dfrac{1}{-|\mathbf{d},\mathbf{u},\mathbf{v}|}\begin{bmatrix}+|\mathbf{b},&\mathbf{u},&\mathbf{v}|\\-|\mathbf{d},&\mathbf{b},&\mathbf{v}|\\-|\mathbf{d},&\mathbf{u},&\mathbf{b}|\end{bmatrix}
\end{equation}

To be able to exploit the similarity of the parallel planes we introduce $\mathbf{w}:=\mathbf{u}\times \mathbf{v}$ and $\mathbf{e}:=\mathbf{d}\times \mathbf{b}$ and rewrite the determinants in equation \refeq{eq:det} as the following:

\begin{equation} \label{eq:intersection}
	\begin{bmatrix} r\\ s\\ t \end{bmatrix}=\dfrac{1}{-\mathbf{d}^T\mathbf{w}}\begin{bmatrix}+\mathbf{b}^T\mathbf{w}\\-\mathbf{e}^T\mathbf{v}\\+\mathbf{e}^T\mathbf{u}\end{bmatrix}
\end{equation}

At this point, we lay out the Algorithm \ref{alg:LineNetworkSampling} that iterates over the dimensions, intersection-planes, and lines to exploit their similarity.
Accordingly, in order, we compute and check $\mathbf{w}, \mathbf{\delta}, \mathbf{b}, r, \mathbf{e}, s, t$ to ensure that we are not wasting computation if the intersection is outside of the boundaries of line-segment and frame.

\begin{table}[H]
	\footnotesize
	\captionsetup{labelformat=empty}
	\caption{\textbf{Algorithm \ref{alg:LineNetworkSampling}}: Line Network Sampling}
	\centering
	\begin{tabularx}{\textwidth}{
    >{\raggedleft\arraybackslash\hsize=0.25\hsize}X
    >{\raggedright\arraybackslash\hsize=0.35\hsize}X
    >{\raggedright\arraybackslash\hsize=2.4\hsize}X
} 
    \toprule
    \textbf{Input} 
    & \textbf{Data Type} 
    & \textbf{Input Name: Notes} 
    \\ 
    \toprule
    $\mathcal{L}$ 
    & Line Set
    & Including vertices and edges 
    \\ 
    $\mathcal{P}$ \comment{@pirouz, @shervin, currently the plane is not used in the algorithm, should we generalize the algorithm to take the plane as the input? or should we remove the plane from the inputs and only stick to voxelizations parallel to the principal axes?}
    & Plane in $\mathbb{R}^3$ 
    & An oriented plane in $\mathbb{R}^3$consisting of 1+2 vectors, respectively indicating the origin, the X-axis, and the Y-axis of the plane, with the default value as the global XY plane. 
    \\ 
    $\boldsymbol{\sigma}_{3\times1}$ 
    & Vector of float 
    & Unit Size Vector: a vector whose components represent the desired voxel size in X, Y, and Z directions 
    \\ 
    \toprule
    \textbf{Output} 
    & \textbf{Data Type} 
    & \textbf{Output Name: Notes} 
    \\ 
    \toprule
    $\mathbf{X}$ 
    & Array of Points
    & list of coordinates of voxel points in the sample point cloud $\in \mathbb{R}^3$
    \\ 
    \toprule
    \multicolumn{3}{p{0.97\textwidth}}{
        $\textbf{Problem}$: Given a line-set $\mathcal{L}$ with vertices in $\mathbb{R}^3$, which is oriented in plane $\mathcal{P}$, and a vector of unit sizes $\boldsymbol{\sigma}$, it is desired to find a set of points $\mathbf{X}$ in $\mathbb{R}^3$, which contains the topological structure of the mesh.
    } 
    \\ 
\end{tabularx}
	\label{tab:LineNetworkSampling}

	\begin{algorithm*}[H]
		\footnotesize
		\setstretch{1.05}
		\caption{Line Network Sampling}
		\label{alg:LineNetworkSampling}
		\SetKwProg{Procedure}{LineNetworkSampling}{:}{}
	\Procedure{($\mathcal{L}$, $\mathcal{P}$ , $\boldsymbol{\sigma}$)}{
		$\mathbf{B}_\mathcal{L}$ = $\nint{BoundingBox(\mathcal{L})\oslash \boldsymbol{\sigma}}$ 
		\tcp{dividing the bounding box of the mesh in $\mathbb{R}^3$ based on the  $\mathcal{P}$ frame by $\boldsymbol{\sigma}$ element-wise, and then rounding it to nearest integer $\in \mathbb{Z}^3: [[i_{min}, j_{min}, k_{min}],[i_{max}, j_{max}, k_{max}]]$}
		$[m, n, o] = \mathbf{B}_\mathcal{L}[1] - \mathbf{B}_\mathcal{L}[0]$ \\
		initialize $\mathbf{X}=[]$ \\
		\For{each axis $a_n\in \{0,1,2\}$}{
			$a_r=(a_n+1)\%3$\tcp{identify the right axis}
			$a_f=(a_n+2)\%3$\tcp{identify the front axis} 
			$\mathbf{u}=\text{diag}([m, n, o]^T)\boldsymbol{\sigma}[a_r,:]$ \\
			$\mathbf{v}=\text{diag}([m, n, o]^T)\boldsymbol{\sigma}[a_f,:]$ \\
			$\mathbf{w}:=\mathbf{u}\times \mathbf{v}$ 
			\tcp{compute $\mathbf{w}$ for all the planes along the current axis $a_n$}
			\For{each plane $\mathpzc{p}=(\mathbf{c}_k, \mathbf{u}, \mathbf{v})$ prependicular to $a_n$; enumerated by $k \in \{1, ..., [m, n, o]^T[a_n]\}$}{ 
				$\mathbf{c}_k[a_n] = \mathbf{B}_\mathcal{L}[a_n,0]+ (k - 0.5)\boldsymbol{\sigma}[a_n]$\\
				$\mathbf{c}_k[a_r] = \mathbf{B}_\mathcal{L}[a_r,0]$\\
				$\mathbf{c}_k[a_f] = \mathbf{B}_\mathcal{L}[a_f,0]$ \\
				\For{ each line $\mathpzc{l}=(\mathbf{p}, \mathbf{d})$ in $\mathcal{L}$}{
					$\delta:=-\mathbf{d}^T\mathbf{w}$ 
					\tcp{compute the detereminant}
					\lIf{$\delta\in[-\epsilon,+\epsilon]$}{continue with next line}
					$\mathbf{b}:=\mathbf{p} - \mathbf{c}$ \\
					\lIf{$r := \frac{\mathbf{b}^T\mathbf{w}}{\delta} \notin [0, 1]$}{continue with next line} 
					$\mathbf{e}:=\mathbf{d}\times \mathbf{b}$ \\
					\lIf{$s := -\frac{\mathbf{e}^T\mathbf{v}}{\delta} \notin [0,1]$}{continue with next line} 
					\lIf{$t := +\frac{\mathbf{e}^T\mathbf{u}}{\delta} \notin [0,1]$}{continue with next line} 
					$\mathbf{x}:=\mathbf{p}+r\mathbf{d}$\\
					append $\mathbf{x}$ to $\mathbf{X}$ 
				}
			} 
		}
	}
	\end{algorithm*}
\end{table}

\subsubsection{Mesh Surface Sampling}
\begin{figure}
	\centering
	\begin{minipage}{.5\textwidth}
	  \centering
	  \includegraphics[trim={13.5cm 6cm 8.5cm 5.5cm},clip,width=\textwidth]{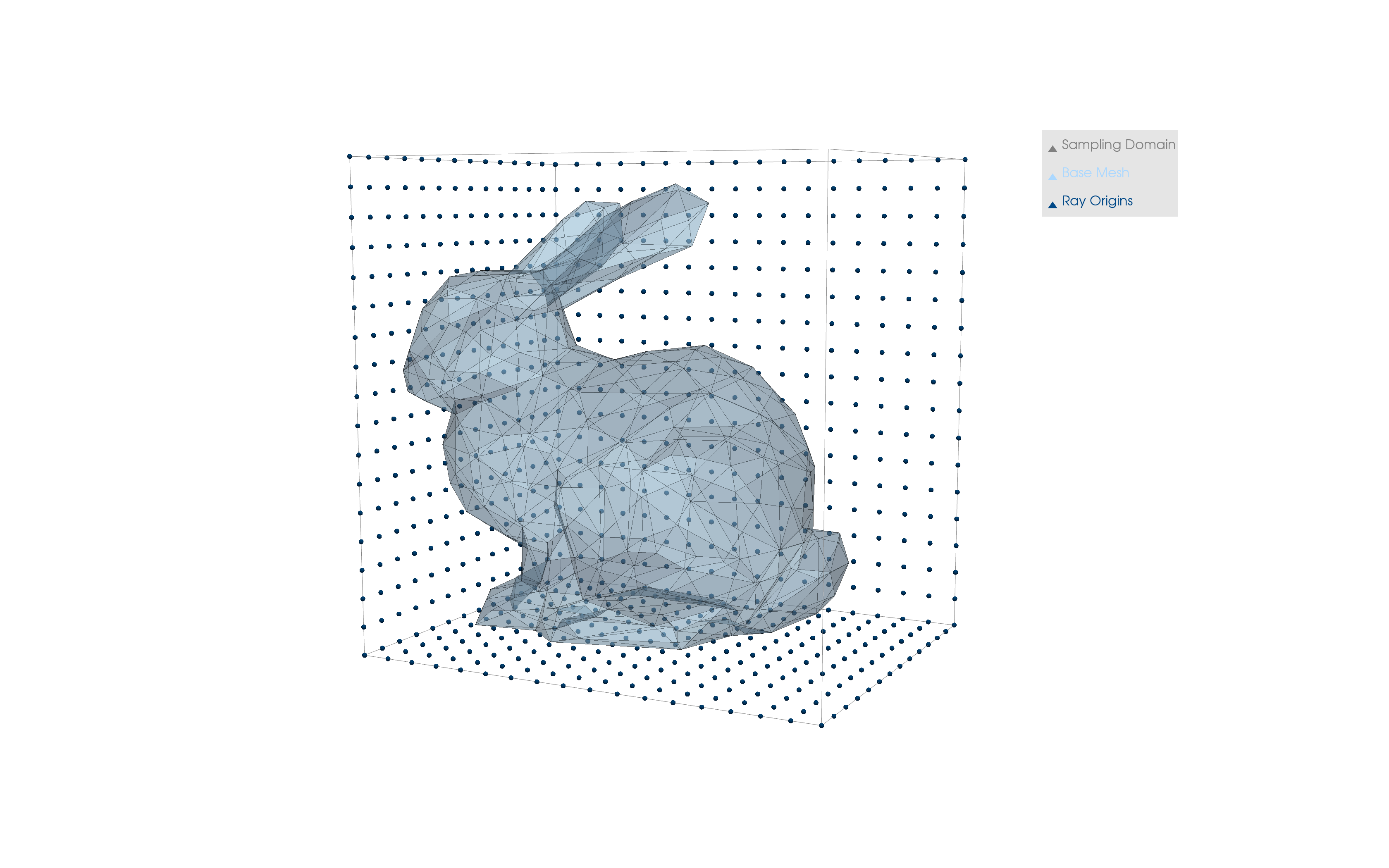}
	  \captionof{figure}{Ray Oigins}
	  \label{fig:rayorigins}
	\end{minipage}%
	\begin{minipage}{.5\textwidth}
	  \centering
	  \includegraphics[trim={13.5cm 6cm 8.5cm 5.5cm},clip,width=\textwidth]{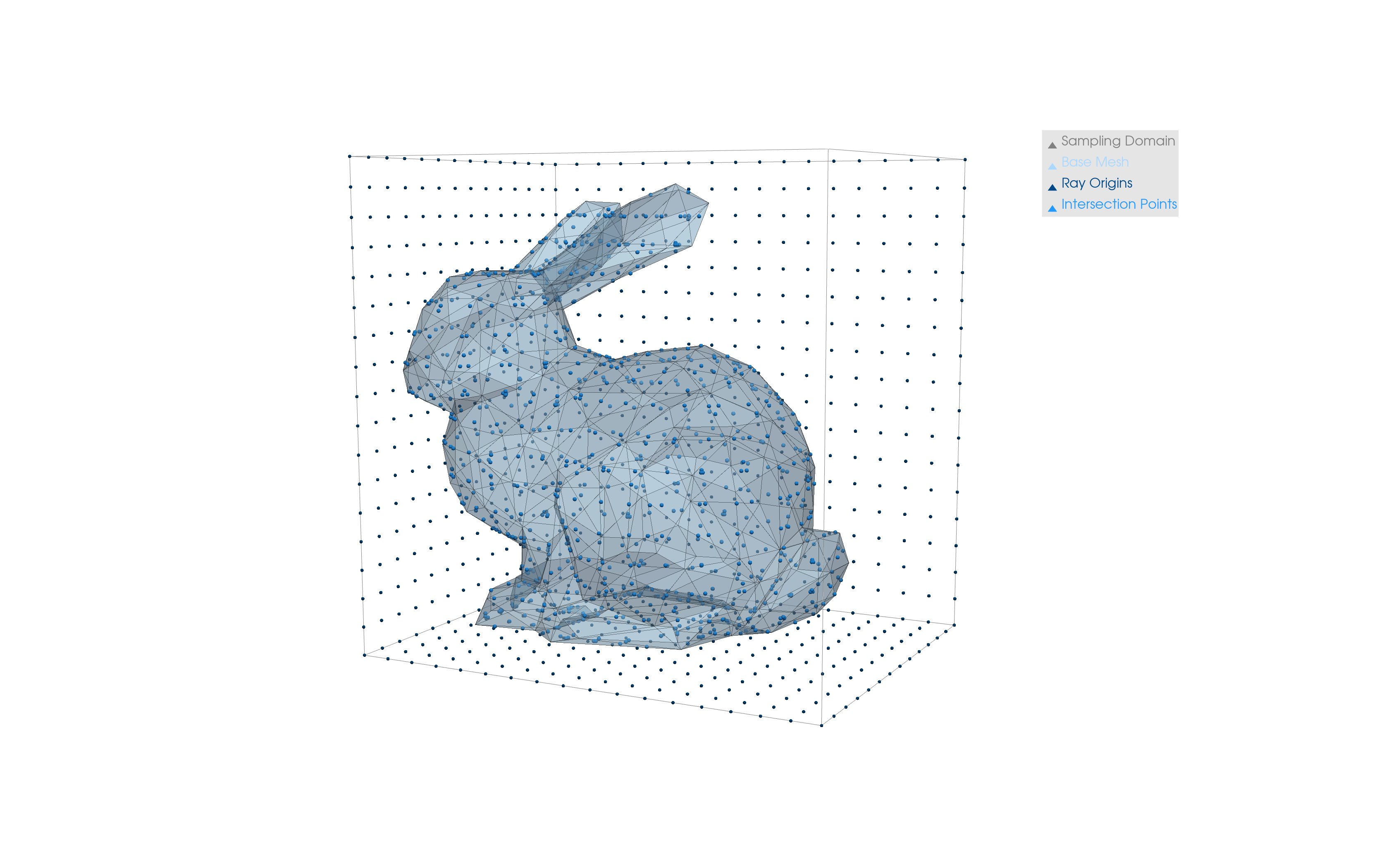}
	  \captionof{figure}{Sampled Point Cloud}
	  \label{fig:sampledpointcloud}
	\end{minipage}
\end{figure}

The second algorithm describes a Mesh Surface Sampling process using lines as the intersection object.
We utilize the mathematical extension of the Ray-Triangle intersection in the previous section (see equation \ref{eq:intersection}) to exploit the regularity of the voxel grid.
We need to construct the intersection objects in accordance with the simplest form of connectivity type (see Table \ref{tab:ConnectivityType}) in the stencil that the Graph Construction (Algorithm \ref{alg:GraphConstruction}) uses further down the pipeline.
In the algorithm below, we utilize the \emph{Conservative voxelization}.
Thus the intersection lines are aligned with voxel boundaries; in an example grid of $m$ by $n$ by $o$ voxels, respectively aligned with the $X$, $Y$, and $Z$ axes.
Consequently, we need $(n+1)\times(o+1)$, $(o+1)\times(m+1)$, and $(m+1)\times(n+1)$ intersection-lines or rays for sampling the input surface mesh within such a bounding grid.
Figure \ref{fig:rayorigins} visualizes the ray origins and Figure \ref{fig:sampledpointcloud} indicates the extracted sampling of the mesh surface as a point cloud.

Utilizing the algebraic formulation of intersection in equation \ref{eq:intersection}, we lay out the Algorithm \ref{alg:MeshSurfaceSampling} that iterates over the dimensions, triangles, and intersection-lines to exploit their similarity.
Accordingly, in order, we compute and check $\mathbf{w}, \mathbf{\delta}, \mathbf{b}, r, \mathbf{e}, s, t$ to ensure that we are not wasting computation if the intersection is outside of the boundaries of line and triangle.

\begin{table}[H]
	\footnotesize
	\captionsetup{labelformat=empty}
	\caption{\textbf{Algorithm ~\ref{alg:MeshSurfaceSampling}}: Mesh Surface Sampling}
	\centering
    \begin{tabularx}{\textwidth}{
    >{\raggedleft\arraybackslash\hsize=0.25\hsize}X
    >{\raggedright\arraybackslash\hsize=0.35\hsize}X
    >{\raggedright\arraybackslash\hsize=2.4\hsize}X
} 
    \toprule
    \textbf{Input} 
    & \textbf{Data Type} 
    & \textbf{Input Name: Notes} 
    \\ 
    \toprule
    $\mathcal{M}$$\rm{(V,F)}$ 
    & Mesh 
    & Including faces and vertices: A triangulated (2-manifold)surface mesh (a.k.a. a Triangular Irregular Network).
    \\ 
    $\mathcal{P}$ \comment{@pirouz, @shervin, currently the plane is not used in the algorithm, should we generalize the algorithm to take the plane as the input? or should we remove the plane from the inputs and only stick to voxelizations parallel to the principal axes?}
    & Plane in $\mathbb{R}^3$ 
    & An oriented plane in $\mathbb{R}^3$consisting of 1+2 vectors, respectively indicating the origin, the X-axis, and the Y-axis of the plane, with the default value as the global XY plane. 
    \\ 
    $[\boldsymbol{\sigma}]_{3\times1}$ 
    & Vector of float 
    & Unit Size Vector: a vector whose components represent the desired voxel size in X, Y, and Z directions 
    \\ 
    \toprule
    \textbf{Output} 
    & \textbf{Data Type} 
    & \textbf{Output Name: Notes} 
    \\ 
    \toprule
    $\mathbf{Q}$ 
    & Array of Points
    & list of coordinates of points in the point cloud $\in \mathbb{R}^3$
    \\ 
    \toprule
    \multicolumn{3}{p{0.97\textwidth}}{
        $\textbf{Problem}$: Given a surface as a triangular mesh $\mathcal{M}$ with vertices in $\mathbb{R}^3$, which is oriented in plane $\mathcal{P}$, and a vector of unit sizes $\mathbf{\sigma}$, it is desired to find a set of points $\mathbf{Q}$ in $\mathbb{R}^3$, which contain the topological structure of the mesh.
    } 
    \\ 
\end{tabularx}
	\label{tab:MeshSurfaceSampling}

	\begin{algorithm*}[H]
		\footnotesize
		\setstretch{1.05}
		\caption{Mesh Surface Sampling Algorithm}
		\label{alg:MeshSurfaceSampling}
		\SetKwProg{Procedure}{MeshSurfaceSampling}{:}{}
\Procedure{( $\mathcal{M}$, $\mathcal{P}$ , $\mathbf{\sigma}$)}{
	$\mathbf{B}_\mathcal{M}$ = $\nint{BoundingBox(\mathcal{M})\oslash \mathbf{\sigma}}$ 
	\tcp{dividing the bounding box of the mesh in $\mathbb{R}^3$ based on the  $\mathcal{P}$ frame by $\mathbf{\sigma}$ element-wise, and then rounding it to nearest integer $\in \mathbb{Z}^3: [[i_{min}, j_{min}, k_{min}],[i_{max}, j_{max}, k_{max}]]$}
	$[m, n, o] = \mathbf{B}_\mathcal{M}[1] - \mathbf{B}_\mathcal{M}[0]$ \\
	initiate $\mathbf{Q}$ \\

	\For{ each triangle $=(\mathbf{x_0, x_1, x_2}) \in \mathcal{M}$}{
		$\mathbf{u} = \mathbf{x_1 - x_0}; \mathbf{v} = \mathbf{x_2 - x_0}; \mathbf{c} =  \mathbf{x_0}$; \\
		$\mathbf{w}:=\mathbf{u}\times \mathbf{v}$ \\

		\For{each axis $a_m \in \{0,1,2\}$}{
			$a_r=(a_m+1)\%3$ ; $a_f=(a_m+2)\%3$ 
			\tcp{identify right and front axis}
			$\mathbf{d}=\text{diag}([m, n, o]^T)[a_m,:]$ \\
			$\delta:=-\mathbf{d}^T\mathbf{w}$ 
			\tcp{compute the detereminant}
			\lIf{$\delta\in[-\epsilon,+\epsilon]$}{continue with next axis}
			\For{each voxel row in $a_r$; enumerated by $k_0 \in \{0, 1, ..., [m, n, o]^T[a_r]\}$}{ 

				\For{each voxel column in $a_f$; enumerated by $k_1 \in \{0, 1, ..., [m, n, o]^T[a_f]\}$}{ 
				$\mathbf{p}[a_m] = \mathbf{B}_\mathcal{M}[a_m,0]$; $\mathbf{c}[a_r] = k_0 - 0.5$; $\mathbf{c}[a_f] = k_1 - 0.5$ \\

				$\mathbf{b}:=\mathbf{p} - \mathbf{c}$ \\
				\lIf{$r = \frac{\mathbf{b}^T\mathbf{w}}{\delta} \notin [r_{min}, r_{max}]$}{continue with next line} 
				$\mathbf{e}:=\mathbf{d}\times \mathbf{b}$ \\
				\lIf{$s = -\frac{\mathbf{e}^T\mathbf{v}}{\delta} \notin [s_{min}, s_{max}]$}{continue with next line} 
				\lIf{$t = \frac{\mathbf{e}^T\mathbf{u}}{\delta} \notin [t_{min}, t_{max}]$}{continue with next line} 
				$\mathbf{q}[a_m] = r$; $\mathbf{q}[a_r] = \mathbf{c}[a_r]$; $\mathbf{q}[a_f] = \mathbf{c}[a_f]$ \\
				add $\mathbf{q}$ to $\mathbf{Q}$ 
				}
			}
		} 
	}
}

	\end{algorithm*}
\end{table}

\subsubsection{Volumetric Mesh Sampling}

The volumetric meshes (3D) embedded in $\mathbb{R}^3$, are mainly represented by the surface mesh that represents their boundary, with the condition that the boundary mesh is closed.
Therefore, we adjust the Algorithm \ref{alg:MeshSurfaceSampling} to sample boundary mesh and construct intervals on each intersection-line.
Next, we discretize the 1D intervals to create samples within the volume.

More specifically, we alter the order of the iterations in algorithm \ref{alg:VolumetricMeshSampling} to the axis, intersection-line, and triangles as we need to construct the intervals on the line-basis.
The core idea is that based on the winding number, we can assume that a closed boundary would always intersect with a ray an even number of times, given that the ray extends across the bounding box.
Therefore, we maintain that pairs of intersection points create intervals on the line corresponding to a one-dimensional sample of the volumetric mesh.
Therefore, by discretizing these intervals, we will generate enough samples to reconstruct the volumetric mesh's interior in the graph construction later.

\begin{table}[!htb]
	\footnotesize
	\captionsetup{labelformat=empty}
	\caption{\textbf{Algorithm ~\ref{alg:VolumetricMeshSampling}}: Volumetric Mesh Sampling}
	\centering
    \begin{tabularx}{\textwidth}{
    >{\raggedleft\arraybackslash\hsize=0.25\hsize}X
    >{\raggedright\arraybackslash\hsize=0.35\hsize}X
    >{\raggedright\arraybackslash\hsize=2.4\hsize}X
} 
    \toprule
    \textbf{Input} 
    & \textbf{Data Type} 
    & \textbf{Input Name: Notes} 
    \\ 
    \toprule
    $\mathcal{M}$$\rm{(V,F)}$ 
    & Mesh 
    & Including faces and vertices representing the boundary of a volumetric mesh.
    \\ 
    $\mathcal{P}$ \comment{@pirouz, @shervin, currently the plane is not used in the algorithm, should we generalize the algorithm to take the plane as the input? or should we remove the plane from the inputs and only stick to voxelizations parallel to the principal axes?}
    & Plane in $\mathbb{R}^3$ 
    & An oriented plane in $\mathbb{R}^3$consisting of 1+2 vectors, respectively indicating the origin, the X-axis, and the Y-axis of the plane, with the default value as the global XY plane. 
    \\ 
    $[\boldsymbol{\sigma}]_{3\times1}$ 
    & Vector of float 
    & Unit Size Vector: a vector whose components represent the desired voxel size in X, Y, and Z directions 
    \\ 
    \toprule
    \textbf{Output} 
    & \textbf{Data Type} 
    & \textbf{Output Name: Notes} 
    \\ 
    \toprule
    $\mathbf{Q}$ 
    & Array of Points
    & list of coordinates of points in the point cloud $\in \mathbb{R}^3$
    \\ 
    \toprule
    \multicolumn{3}{p{0.97\textwidth}}{
        $\textbf{Problem}$: Given a surface as a triangular mesh $\mathcal{M}$ with vertices in $\mathbb{R}^3$, which is oriented in plane $\mathcal{P}$, and a vector of unit sizes $\mathbf{\sigma}$, it is desired to find a set of points $\mathbf{Q}$ in $\mathbb{R}^3$, which contain the topological structure of the mesh.
    } 
    \\ 
\end{tabularx}
	\label{tab:VolumetricMeshSampling}
	\begin{algorithm*}[H]
		\footnotesize
		\setstretch{1.05}
		\caption{Volumetric Mesh Sampling Algorithm}
		\label{alg:VolumetricMeshSampling}
		\SetKwProg{Procedure}{VolumetricMeshSampling}{:}{}
\Procedure{( $\mathcal{M}$, $\mathcal{P}$ , $\mathbf{\sigma}$)}{
	$\mathbf{B}_\mathcal{M}$ = $\nint{BoundingBox(\mathcal{M})\oslash \mathbf{\sigma}}$ 
	\tcp{dividing the bounding box of the mesh in $\mathbb{R}^3$ based on the  $\mathcal{P}$ frame by $\mathbf{\sigma}$ element-wise, and then rounding it to nearest integer $\in \mathbb{Z}^3: [[i_{min}, j_{min}, k_{min}],[i_{max}, j_{max}, k_{max}]]$}
	$[m, n, o] = \mathbf{B}_\mathcal{M}[1] - \mathbf{B}_\mathcal{M}[0]$ \\
	initiate $\mathbf{Q}$ \\
	
	\For{each axis $a_m \in \{0,1,2\}$}{
		
		$a_r=(a_m+1)\%3$ ; $a_f=(a_m+2)\%3$ 
		\tcp{identify right and front axis}
		$\mathbf{d}=\text{diag}([m, n, o]^T)[a_m,:]$ \\
		\For{each voxel row in $a_r$; enumerated by $k_0 \in \{0, 1, ..., [m, n, o]^T[a_r]\}$}{ 

			initiate $\mathbf{r}$ \\
			\For{each voxel column in $a_f$; enumerated by $k_1 \in \{0, 1, ..., [m, n, o]^T[a_f]\}$}{ 

				\For{ each triangle $=(\mathbf{x_0, x_1, x_2}) \in \mathcal{M}$}{
				$\mathbf{u} = \mathbf{x_1 - x_0}; \mathbf{v} = \mathbf{x_2 - x_0}; \mathbf{c} =  \mathbf{x_0}$; \\
				$\mathbf{w}:=\mathbf{u}\times \mathbf{v}$ \\

				$\delta:=-\mathbf{d}^T\mathbf{w}$ 
				\tcp{compute the detereminant}
				\lIf{$\delta\in[-\epsilon,+\epsilon]$}{continue with next triangle}
				$\mathbf{p}[a_m] = \mathbf{B}_\mathcal{M}[a_m,0]$; $\mathbf{c}[a_r] = k_0 - 0.5$; $\mathbf{c}[a_f] = k_1 - 0.5$ \\

				$\mathbf{b}:=\mathbf{p} - \mathbf{c}$ \\
				$\mathbf{e}:=\mathbf{d}\times \mathbf{b}$ \\
				\If{$r = \frac{\mathbf{b}^T\mathbf{w}}{\delta} \in [r_{min}, r_{max}]$ and $s = -\frac{\mathbf{e}^T\mathbf{v}}{\delta} \in [s_{min}, s_{max}]$ and $t = \frac{\mathbf{e}^T\mathbf{u}}{\delta} \in [t_{min}, t_{max}]$}{
					add $r$ to $\mathbf{r}$ 
				} 
				}
				}
				\lIf{$len(\mathbf{r}) \% 2$ != $0$}{\Return{the boundary is not closed}}
				\For{$k_m \in \{0, 1, ..., [m, n, o]^T[a_m]\}$}{
					\If{$ Sum(k_m > \mathbf{r})\% 2$ != $0$}{
					$\mathbf{q}[a_m] = k_m$; $\mathbf{q}[a_r] = \mathbf{c}[a_r]$; $\mathbf{q}[a_f] = \mathbf{c}[a_f]$ \\
					add $\mathbf{q}$ to $\mathbf{Q}$
					
				}

			}
		} 
	}
}
	\end{algorithm*}
\end{table}

\subsection{Point Cloud Voxelization}\label{sec:voxelization}

\begin{figure}
	\centering
	\includegraphics[width=\textwidth]{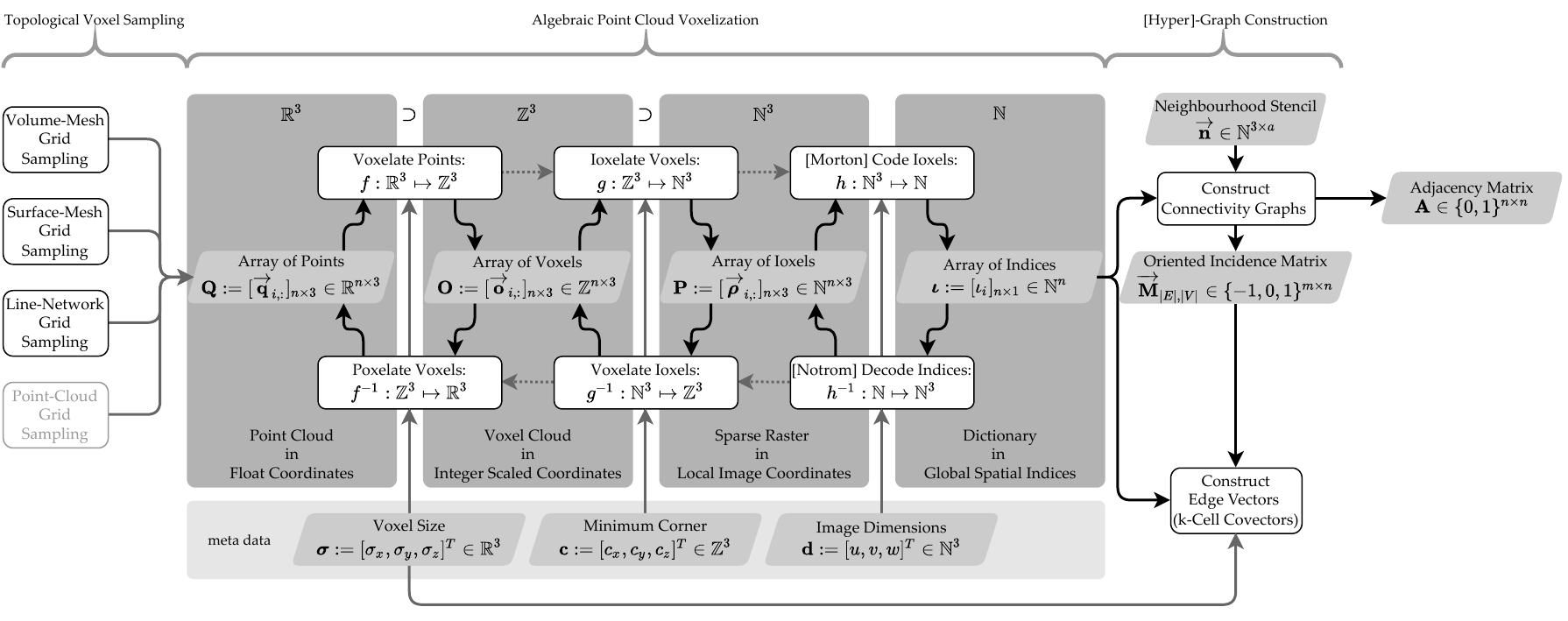}
	\caption{the first part of the proposed workflow: Digital Geometry Sampling and Topological Voxelization}
	\label{fig:sampling_voxelization}
	\vspace{-4mm}
\end{figure}

Irregular point clouds (e.g. LIDAR images) represent samples in the $\mathbb{R}^3$ without any inherent topological structure that specifies the relation of sample points.
However, if these point clouds are the output of the mesh sampling proposed earlier, the topological properties of the original mesh can be retrieved from them.
Nevertheless, this retrieval is dependent on using the same voxel unit size $\boldsymbol{\sigma}$ and plane $\mathcal{P}$ for the Algorithm \ref{alg:PointCloudVoxelization}.
This algorithm is comprised of three smaller steps: Voxelation, Shifting(Ioxelation), and Encoding (see Figure \ref{fig:sampling_voxelization}).

Voxelization ($f$) on its own is a function that takes in the voxel unit size and maps the points from $\mathbb{R}^3$ to $\mathbb{Z}^3$.
This step is the core of the discretization process.
It is important to note that the reverse function of voxelization $f^{-1}$ does not reproduce the original points, as the remainder of the coordinates values that are smaller than the voxel size $\boldsymbol{\sigma}$ is lost:

\begin{align}
	Voxelate\; Points: f: \mathbb{R}^3\mapsto \mathbb{Z}^3
	& \qquad
	\mathbf{v}:= f(\mathbf{p})=\lfloor\mathbf{p}\oslash \boldsymbol{\sigma}\rceil
	\\
	Poxelate\; Voxels: f^{-1}: \mathbb{Z}^3\mapsto \mathbb{R}^3
	& \qquad
	\mathbf{p}:= f^{-1}(\mathbf{v})=\mathbf{v}\odot \boldsymbol{\sigma}
\end{align}

The next step is to shift the voxels into the first quadrant so their coordinates in $\mathbb{Z}^3$ maps to $\mathbb{N}^3$ and can be regarded as three-dimensional indices: Ioxelate ($g$).
The necessary input for this function is the minimum corner of the voxel cloud ($\mathbf{c}$).

\begin{align}
	Ioxelate\; Voxels: g: \mathbb{Z}^3\mapsto \mathbb{N}^3
	& \qquad
	\boldsymbol{\rho}:= g(\mathbf{v})=\mathbf{v}-\mathbf{c}
	\\
	Voxelate\; Ioxels: g^{-1}: \mathbb{N}^3\mapsto \mathbb{Z}^3
	& \qquad
	\mathbf{v}:= g^{-1}(\boldsymbol{\rho})=\boldsymbol{\rho}+\mathbf{c}
\end{align}

The last step is encoding the three-dimensional indices into Morton code, using the three-dimensional interleaving method.
This would effectively map the $\mathbb{N}^3$ to $\mathbb{N}$ which allows for globally unique indices for each voxel.

\begin{align}
	Encode\; Ioxels: h: \mathbb{N}^3\mapsto \mathbb{N}
	& \qquad
	\iota:=h(\boldsymbol{\rho})=Morton(\boldsymbol{\rho})
	\\
	Decode\; Indices: h: \mathbb{N}\mapsto \mathbb{N}^3
	& \qquad
	\boldsymbol{\rho}:=h^{-1}(\boldsymbol{\iota})=Notrom(\boldsymbol{\iota})
\end{align}

\begin{figure}
	\centering
	\begin{minipage}{.5\textwidth}
	  \centering
	  \includegraphics[trim={13.5cm 6cm 8.5cm 5.5cm},clip,width=\textwidth]{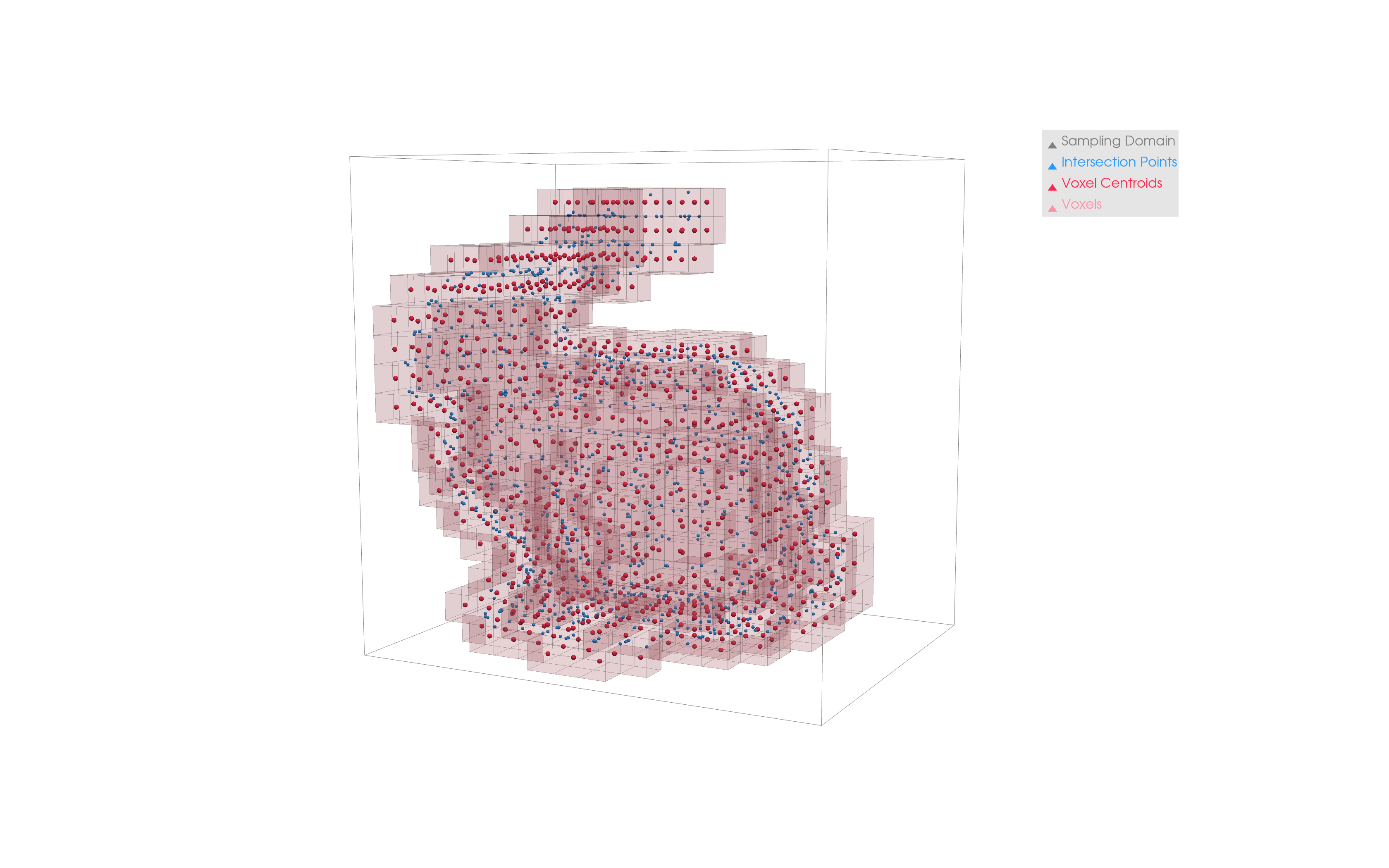}
	  \captionof{figure}{Voxelization}
	  \label{fig:voxelization}
	\end{minipage}%
	\begin{minipage}{.5\textwidth}
	  \centering
	  \includegraphics[trim={13.5cm 6cm 8.5cm 5.5cm},clip,width=\textwidth]{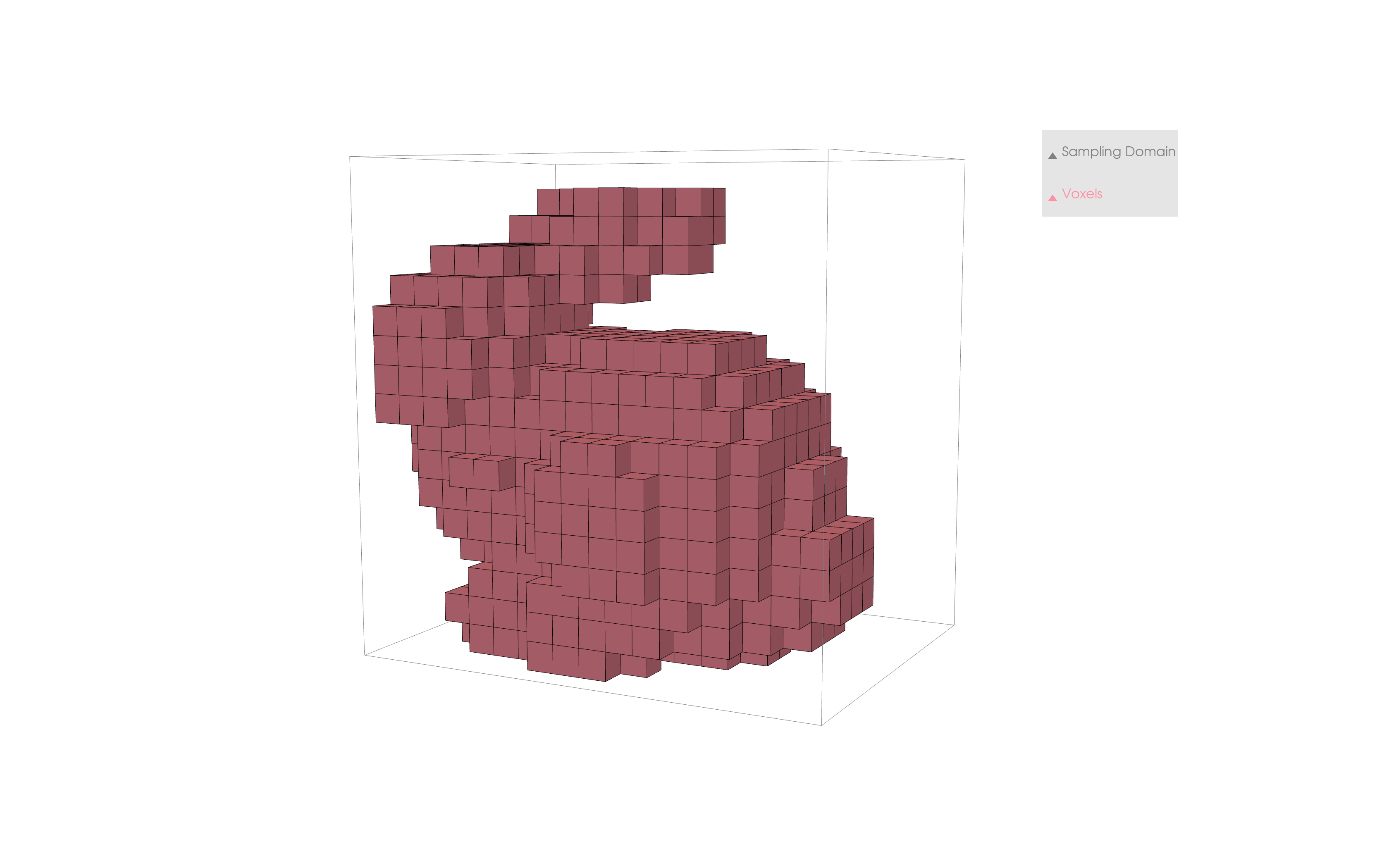}
	  \captionof{figure}{Voxels}
	  \label{fig:voxels}
	\end{minipage}
\end{figure}

Altogether, they start from a point cloud and generate a voxel cloud represented by globally unique spatial Morton indexing.
Figures \ref{fig:voxelization} and \ref{fig:voxels} overlay the point cloud that is the process's input onto the voxel centroids and voxel domains that are the output.

\begin{table}[H]
	\footnotesize
	\captionsetup{labelformat=empty}
	\caption{\textbf{Algorithm ~\ref{alg:PointCloudVoxelization}}: Point Cloud Voxelization}
	\centering
	\begin{tabularx}{\textwidth}{
    >{\raggedleft\arraybackslash\hsize=0.25\hsize}X
    >{\raggedright\arraybackslash\hsize=0.35\hsize}X
    >{\raggedright\arraybackslash\hsize=2.4\hsize}X
} 
    \toprule
    \textbf{Input} 
    & \textbf{Data Type} 
    & \textbf{Input Name: Notes} 
    \\
    \toprule
    $\mathbf{Q}$ 
    & Array of Points
    & list of coordinates of points in the point cloud in $\mathbb{R}^3$
    \\ 
    $\mathcal{P}$ 
    & Plane in $\mathbb{R}^3$ 
    & An oriented plane in $\mathbb{R}^3$consisting of 1+2 vectors, respectively indicating the origin, the X axis, and the Y axis of the plane, with the default value as the global XY plane. 
    \\ 
    $[\mathbf{\sigma}]_{3\times1}$ 
    & Vector of float 
    & Unit Size Vector: a vector whose components represent the desired voxel size in X, Y, and Z directions 
    \\ 
    \toprule
    \textbf{Output} 
    & \textbf{Data Type} 
    & \textbf{Output Name: Notes} 
    \\ 
    \toprule
    $\boldsymbol{\iota}$ 
    & Array of int
    & Array of global spatial morton indices
    \\ 
    \toprule
    \multicolumn{3}{p{0.97\textwidth}}{
        $\textbf{Problem}$: Given an array of points $\mathbf{Q}$ which is oriented in plane $\mathcal{P}$, and a vector of sizes $\boldsymbol{\sigma}$, it is desired a set of morton indices $\boldsymbol{\iota}\subset\mathbb{N}$ as a discrete approximation of the point cloud in question such that the set of voxels $\mathcal{V}$ corresponding to the indices compactly and correctly represents the input point cloud. 
        Correctness must be verifiable in terms of point-set topological properties of the input point-cloud and the output voxel cloud being on a par with one another. 
        \comment{somewhere in the text after or before this we can elaborate on this point and explain that the parity is about a simplicial complex built over the input point cloud (e.g. a Čech complex or a Vietoris–Rips complex) having the same Euler-Poincare characteristic as the simple graph or a more sophisticated complex made upon the output voxel cloud}
    } 
    \\ 
\end{tabularx}
	\label{tab:PointCloudVoxelization}
	\begin{algorithm*}[H]
		\footnotesize
		\setstretch{1.05}
		\caption{Point Cloud Voxelization}
		\label{alg:PointCloudVoxelization}
		\SetKwProg{Procedure}{VoxelatePointCloude}{:}{}
	\Procedure{( $\mathbf{Q}, \mathcal{P}, \mathbf{\sigma}$)}{
		
		initiate $\boldsymbol{\iota}$ \\
		\For{point $\mathbf{x} \in \mathbf{Q}$}{
			$\mathbf{o} := \lfloor \mathbf{x} \oslash \boldsymbol{\sigma} \rceil$ \tcp{voxelate}
			$\boldsymbol{\rho} := \mathbf{o} - \mathcal{P}_c$ \tcp{Ioxelate}
			$\iota := MortonInterleave3D(\boldsymbol{\rho})$ \tcp{generate morton index}
			add $\iota$ to $\boldsymbol{\iota}$ 
		}
	}
	\end{algorithm*}
\end{table}


\subsection{Graph Construction}\label{sec:GraphConstruction}

In this step, given the voxel cloud represented by Morton indices $\boldsymbol{\iota}$, we construct a topological model of their connectivity.
Since the domain space is discrete, we utilize the concept of the stencil to construct a topological model that represents a particular local neighbourhood for each voxel.
Stencils were initially proposed by \cite{emmonsNumericalSolutionPartial1944} for Finite Difference Method operations required for finding numerical solutions to PDEs in regular Cartesian grids (find a recent review in \cite{engwerStencilComputationsPDEbased2017}).
Here we generalize the idea of stencils to extract a topological description of the local connectivity of voxels using bitwise operations on their Morton codes.

The stencils proposed in this paper can be described by a condition array of relative indices $\mathcal{S}_v$ and a relative hyper-edge $\mathcal{S}_e$.
The stencil checks whether a voxel $\mathbf{o}$ of the volume has a specific connectivity pattern described in $\mathcal{S}_v$ and returns a boolean value.
If the boolean value is $TRUE$, we can construct the hyper-edge based on the relative vertex indices of the condition array $\mathcal{S}_v$.
Here is an example of a stencil describing square connectivity in the $YZ$ plane:

\begin{align}
	\mathcal{S}_v :=
	& \Bigl[[0, 0, 0], [0, 1, 0], [0, 1, 1], [0, 0, 1]\Bigr]
	\\
	\mathcal{S}_e :=
	& \Big\{(0, 1), (1, 2), (2, 3), (3, 0)\Big\}
\end{align}

The selected voxel $\mathbf{o}$ is considered as $(0,0)$, and $\mathcal{S}_v$ describes which relative neighbours of $\mathbf{o}$ should be filled for the edges to be constructed.
$\mathcal{S}_e$ describes which edges need to be made within the hyper-edge by their index in $\mathcal{S}_v$.
In this sense, $\mathcal{S}$ can be understood as a graph that functions similar to a kernel in image processing.

To utilize this concept with the Morton index of the voxels, we need to describe the relative neighbours $\mathcal{S}_v$ of a voxel with their Morton index.
Accordingly, the edges between the neighbours $\mathcal{S}_e$ can also be interpreted as a point in a discrete 2D space of the source voxels and destination voxels; the edge space: $\mathbb{N}^2$.
Therefore, the second Morton indexing process allows us to have unique Morton indices for the edges:

\begin{align}
	\mathcal{S}_\iota &:= Morton(\mathcal{S}_v) =
	 \Big\{\texttt{0b000}, \texttt{0b010},\texttt{0b011}, \texttt{0b001}\Big\}
	\\
	\mathcal{S}_\varepsilon &:=
	 \Big\{\texttt{0b001000}, \texttt{0b001110},\texttt{0b000111}, \texttt{0b000001}\Big\}
\end{align}

This indicates that given a voxel index, we can perform a bitwise sum on the interleaved sequence to find the neighbours and check their value.
More specifically, as the voxels are embedded in the $\mathbb{Z}^3$ space, we will utilize the $MortonSum3D$ to find the index of the voxel neighbours based on $\mathcal{S}_\iota$.
When the conditions are checked, we utilize the $MortonSum6D$ to construct the Morton index of the corresponding edges based on $\mathcal{S}_\varepsilon$.



\begin{table}[H]
	\footnotesize
	\captionsetup{labelformat=empty}
	\caption{\textbf{Algorithm ~\ref{alg:GraphConstruction}}: Graph Construction}
	\centering
    \begin{tabularx}{\textwidth}{
    >{\raggedleft\arraybackslash\hsize=0.25\hsize}X
    >{\raggedright\arraybackslash\hsize=0.35\hsize}X
    >{\raggedright\arraybackslash\hsize=2.4\hsize}X
} 
    \toprule
    \textbf{Input} 
    & \textbf{Data Type} 
    & \textbf{Input Name: Notes} 
    \\ 
    \toprule
    $\boldsymbol{\iota}$ 
    & Array of int
    & Array of global spatial morton indices of voxels
    \\ 
    $\mathcal{S}$ 
    & Stencil
    & Stencil is a graph: ($\mathcal{S}_\iota, \mathcal{S}_\varepsilon$).
    \\ 
    \toprule
    \textbf{Output} 
    & \textbf{Data Type} 
    & \textbf{Output Name: Notes} 
    \\ 
    \toprule
    $\mathbf{M}$ 
    & Sparse Matrix
    & Sparse matrix representing the connectivity graph: Vertex-Hyperedge \comment{@pirouz: I think that when the graph is representing a geometrical mesh, it is useful to differentiate between VE, VF, and VB. However, when we are talking about the topological graph (which does not have any geometric attribute, nodes do not have coordinates), we only need a VH(Vertex-Hyperedge) matrix to represent the graph. (maybe even in the case of geometrical mesh VH is sufficient)}
    \\
    $\boldsymbol{\varepsilon}$ 
    & Array of int
    & Array of global spatial morton indices of graph edges
    \\ 
    \toprule
    \multicolumn{3}{p{0.97\textwidth}}{
        $\textbf{Problem}$: Given an array of morton indices of voxels $\boldsymbol{\iota}$ representing volumetric dataset and a graph $\mathcal{S}$ representing stencil that describes the local neighbourhood of interest, find the edge-vertex connectivity matrix $\mathbf{M}_{EV}$ and array of edge morton indices $\boldsymbol{\varepsilon}$  representing the graph that embodies the topological properties of the original mesh.
    } 
    \\ 
\end{tabularx}
	\label{tab:GraphConstruction}

	\begin{algorithm*}[H]
		\footnotesize
		\setstretch{1.05}
		\caption{Graph Construction}
		\label{alg:GraphConstruction}
		\SetKwProg{Procedure}{MortonSum3D}{:}{}
\Procedure{($\iota_0$,$\iota_1$)}{
	$x'$ = \texttt{0b001001} ;
    $y'$ = \texttt{0b010010} ;
    $z'$ = \texttt{0b100100} ;\\
	\Return{$\iota_{sum}$ = (\\
	\qquad (($\iota_0$ \texttt{|} ($z'+y'$)) + ($\iota_1$ \texttt{\&} $x'$) \texttt{\&} $x'$) \texttt{|}\\
	\qquad (($\iota_0$ \texttt{|} ($x'+z'$)) + ($\iota_1$ \texttt{\&} $y'$) \texttt{\&} $y'$) \texttt{|}\\
	\qquad (($\iota_0$ \texttt{|} ($y'+x'$)) + ($\iota_1$ \texttt{\&} $z'$) \texttt{\&} $z'$) )
	}
	}
\SetKwProg{Procedure}{MortonSum6D}{:}{}
\Procedure{($\varepsilon_0$,$\varepsilon_1$)}{
	$x'$ = \texttt{0b000001000001};
    $y'$ = \texttt{0b000010000010};
    $z'$ = \texttt{0b000100000100}; \\
    $a'$ = \texttt{0b001000001000}; 
    $b'$ = \texttt{0b010000010000};
    $c'$ = \texttt{0b100000100000}; \\
    $f'$ = \texttt{0b111111111111}; \\
	\Return{$\varepsilon_{sum}$ = (\\
	\qquad (($\varepsilon_0$ \texttt{|} ($f'-x'$)) + ($\varepsilon_1$ \texttt{\&} $x'$) \texttt{\&} $x'$) \texttt{|} \\
	\qquad (($\varepsilon_0$ \texttt{|} ($f'-y'$)) + ($\varepsilon_1$ \texttt{\&} $y'$) \texttt{\&} $y'$) \texttt{|} \\ 
	\qquad (($\varepsilon_0$ \texttt{|} ($f'-z'$)) + ($\varepsilon_1$ \texttt{\&} $z'$) \texttt{\&} $z'$) \texttt{|} \\ 
	\qquad (($\varepsilon_0$ \texttt{|} ($f'-a'$)) + ($\varepsilon_1$ \texttt{\&} $a'$) \texttt{\&} $a'$) \texttt{|} \\ 
	\qquad (($\varepsilon_0$ \texttt{|} ($f'-b'$)) + ($\varepsilon_1$ \texttt{\&} $b'$) \texttt{\&} $b'$) \texttt{|} \\ 
	\qquad (($\varepsilon_0$ \texttt{|} ($f'-c'$)) + ($\varepsilon_1$ \texttt{\&} $c'$) \texttt{\&} $c'$) )
	}
	}	
\SetKwProg{Procedure}{GraphConstruction}{:}{}
\Procedure{($\boldsymbol{\iota}$, $\mathcal{S}$ )}{
	initiate $\boldsymbol{\varepsilon}, \mathbf{M}$ \\
	\For{$\iota \in \boldsymbol{\iota}$}{
		$\boldsymbol{\iota}_{n} := MortonSum3D(\iota, \mathcal{S}_\iota)$ \tcp{find the morton code of neighbours}
		\If{$All(\boldsymbol{\iota}_{neighbors})$}{
			$\varepsilon_0 = MortonInterleave2D(\iota, \iota)$ \tcp{interleave the edge index for $(\iota, \iota)$}
			$\boldsymbol{\varepsilon}_\iota = MortonSum6D(\varepsilon_0, \mathcal{S}_\varepsilon)$ \tcp{sum the morton index of $\varepsilon_0$ with the stencil's edges $\mathcal{S}_\varepsilon$ to find the new edges index}
			add $\boldsymbol{\varepsilon}_\iota$ to $\boldsymbol{\varepsilon}$ \\
			add $\boldsymbol{\varepsilon}_\iota$ to $\mathbf{M}$
		}

	}
	}	
	\end{algorithm*}
\end{table}

The concept of stencil as a graph can easily be generalized to hyper-graph, i.e. graph containing hyper-edges that can correspond to faces or cells.
In those cases, the hyper-edge can function as the topological object and the Morton indexing can also be generalized to generate globally unique indices for hyper-edges.
As an example, algorithms like Marching Cube or Marching Square that primarily rely on topological objects like cubes and squares can be redefined based on this process.

\subsection{Derivation of Differential Operators}\label{sec:DiscreteVectorCalculus}
In Machine Learning on Graphs and Signal Processing on graphs or meshes, and especially in Scientific Computing and Computer Aided Engineering for  Simulations involving PDE or the so-called Geosimulations in geospatial sciences it is common to represent functions of space and time NOT in analytical closed forms for mapping input Cartesian coordinates to output values but as \emph{sampled} fields represented with a finite set of samples, i.e. scalar or vector values attributed to some sample points in a spatial domain, which may be meshed or connected through a network or just form a point cloud with attributes. For a discrete function defined as a vector of float attributed generally to the vertices of a graph, or specifically to the voxels of a voxel graph $\Gamma=(V,E), E\subset V\times V, n:=|V|, m:=|E|$ we can form a vectore in the form of $\mathbf{f}=[f_i]_{n\times 1}$.
In such settings, one typically requires to compute such things as spatial integrals (line/curve integrals, surface integrals, or volume integrals) and spatial partial derivatives (for computing such things as the gradient, divergence, curl, and Laplacian).
Then, from a computational mathematics point of view, it is desirable to have differential and integral operators in the form of matrices that can be multiplied from the left to such vectors.
By such discrete operators one can have an inherently discrete version of operations defined in vector calculus such as gradient, divergence, Laplacian differential operators, and even a Reimann integral operator.
Without diving into the details of the Cartesian coordinates or other geometric details for that matter, here we propose a topologically general idea for representing partial derivatives and differential forms (integrands) within the higher-dimensional in-between spaces of k-cells connecting the vertices.
Without loss of generality for higher-dimensional topological cell complexes or even irregular meshes, here we briefly introduce a family of operators for graphs constructed out of voxels.

We present two new mathematical results: one of which is a new formulation of the discrete Laplacian, derived from two exactly dual discrete operators (gradient and divergence) that are generally applicable to graphs of any kind, regardless whether they are obtained from voxels or not; the other one being a discrete Reimanian integral operator in the form of a covector (1-form or row vector).
Remarkably, these results are derived thanks to the characterization of the edge-space of the graphs as the natural place for defining partial differentials and Reimann integrands.

The key idea here is to utilize oriented adjacency matrices as introduced in Table\ref{tab:CombinatorialGraphs} for defining differential forms and their unoriented versions for defining spatial integrals. Here we show a chain of derivations all of which are obtained from the oriented edge-vertex incidence matrix on voxel graphs.

Before moving on to the derivation of the differential/integral operators, it is necessary to also clarify the claim implicitly shown in the Table\ref{tab:CombinatorialGraphs}:
The adjacency matrices indicating undirected or bi-directed relations between cells of the same dimensions can be obtained by matrix multiplication of [sparse] unoriented incidence matrices and the oriented incidence matrices can be multiplied with one another to obtain other incidence matrices (extending the definitions given in \cite{nourian2016spectral},\cite{batty2004new}, and\cite{weilerEdgeBasedDataStructures1985}).
For example the Vertex to Vertex adjacency matrix can be obtained from unoriented Edge to Vertex incidence matrix and its transpose:

\begin{equation}
	\mathbf{A}_{|V|,|V|}=\overline{\mathbf{M}}_{|V|,|E|}\overline{\mathbf{M}}_{|E|,|V|}
\end{equation}
or the Face to Face adjacency matrix can be obtained from Face to Vertex incidence matrices which are by definition unoriented:

\begin{equation}
	\mathbf{A}_{|F|,|F|}=\overline{\mathbf{M}}_{|F|,|V|}\overline{\mathbf{M}}_{|V|,|F|}
\end{equation}
, which can in turn be derived from unoriented Face to Edge and Edge to Vertex incidence matrices:
\begin{equation}
		\mathbf{M}_{|F|,|E|}=\overline{\mathbf{M}}_{|F|,|E|}\overline{\mathbf{M}}_{|E|,|V|}
\end{equation}
Of course, these matrices will almost always be sparse and so they should practically all be represented as sparse matrices.
One particularly interesting example is the Oriented Edge to Vertex incidence matrix, from which we obtain our new results (a new Laplacian operator and spatial integral operators):


\begin{equation}
	\overrightarrow{\mathbf{M}}_{|E|,|V|}\in \{-1,0,1\}^{m\times n}
\end{equation}
, whose vertices and edges in a voxel graph are indexed with 3D and 6D Morton Indices (respectively denoted as $\iota$ and $\varepsilon$) as proposed above (with a simplified notation for the sake of brevity):
\begin{equation}
	\mathbf{M}:=\overrightarrow{\mathbf{M}}_{|E|,|V|}=[M_{\varepsilon,\iota}]_{m\times n}
\end{equation}
, in which the oriented Edge to Vertex incidence entries are defined as:
\begin{equation}
	M_{\varepsilon,\iota}=\left\{\begin{matrix}
		-1 & if \; \varepsilon=(\iota,t), \forall t \in V\\
		+1 & if \; \varepsilon=(s,\iota), \forall s \in V\\
		0  & otherwise
		\end{matrix}\right..
\end{equation}
From this matrix we derive the differential operators and from its unoriented version defined below we derive barrycentres with which we can define Riemannian integral operators for discrete line/curve integrals, surface integrals or volume integrals.
\begin{equation}
	\overline{\mathbf{M}}=\text{abs}\left(\overrightarrow{\mathbf{M}}_{|E|,|V|} \right)
\end{equation}
For defining the differential operators that are by definition oriented and as proposed aligned with the edges, we shall need edge vectors, which can remarkably be obtained from the same Oriented Edge to Vertex incidence matrix.
To do so, we firstly need to define a matrix containing vertex coordinates as below:
\begin{equation}
	\mathbf{V}:=[\overrightarrow{\mathbf{o}}_{\iota,:}]_{n,3}
\end{equation}
, and so, the edge vectors can be obtained in one go, i.e. algebraically, as below:
\begin{equation}
	\mathbf{E}=[\overrightarrow{\mathbf{e}}_{\varepsilon, :}]_{m\times 3}=\mathbf{M}\mathbf{V}
\end{equation}
; using which the edge lengths can be found as the squared 2-norms of the edge vectors:
\begin{equation}
	\boldsymbol{\xi}^2=[\xi^2_e]_{m\times 1}=\text{diagonal}(\mathbf{E}\mathbf{E}^T).
\end{equation}
\begin{equation}
	\boldsymbol{\xi}=[\xi_e]_{m\times 1}=\text{sqrt}\left(\text{diagonal}(\mathbf{E}\mathbf{E}^T)\right)
\end{equation}
Now, we can define a diagonal matrix containing the edge lengths, the reciprocals of which can be used to make its inverse that will be used in the definition of the differential operators.
\begin{equation}
	\boldsymbol{\Xi}=[\xi_{e,e}]_{m\times m}=\text{diag}(\boldsymbol{\xi})
\end{equation}
The first fundamental operator of interest is the discrete gradient operator defined over the edge space of the graph.
We propose the following discrete gradient operator as a mapping from the vertex space of the graph to its edge space:
\begin{equation}
	\overline{\nabla}=\mathbf{G}:=\mathbf{\Xi}^{-1}\mathbf{M}\in \mathbb{R}^{m\times n}, \overline{\nabla}: \mathbb{R}^n\mapsto \mathbb{R}^m
\end{equation}
The claim is that this operator accurately and unambiguously approximates the continuous gradient operator:
\begin{equation}
	\text{grad}(f(\mathbf{x}))=\nabla f(\mathbf{x})=\left[\dfrac{\partial f(\mathbf{x})}{\partial x_i}\right]_{d}, \mathbf{x}\in \mathbb{R}^d.
\end{equation}
Nevertheless, the point is that the defenition above is focused on functions that are analytically defined and thus it needs to be computed and evaluated at every point of the space if the analytic defiition of the function is at hand.
However, in situations weher the function is only sampled in space, this is not convenient.
In fact, the Finite Difference Method for discretizing gradients based on this definition also brings about other challenges such as the need for differentiating between Forward Differences, Backward Differences, and Central Differences because of the missing half-spaces of the edge pixels or voxels of images.
On the contrary, our proposed gradient operator conveniently lives in the m-dimensional space of the edges of the graph and so the direction of the edges to which the partial differentials are attributed makes the gradient vectors at every locality without the need to attribute them to unoriented vertices, since the edges are already oriented.
thus we claim the following (omitting similar claims about the divergence and Laplacian operator for brevity):
\begin{equation}
	\text{grad}(f(\mathbf{x}))=\nabla f(\mathbf{x})=\left[\dfrac{\partial f(\mathbf{x})}{\partial x_i}\right]_{d} \bigg|_{\mathbf{x}}, \forall \mathbf{x}\in \Omega \approx \mathbf{G}\mathbf{f}.
\end{equation}
Similarly, by virtue of the duality of the gradient and divergence operators and also according to the Divergence Theorem, we propose the discrete divergence operator as a mapping from the edge space to the vertex space of the graph:
\begin{equation}
	\overline{\nabla}^T=\mathbf{D}:=\mathbf{M}^T\boldsymbol{\Xi}^{-1}\in \mathbb{R}^{n\times m}, \overline{\nabla}^T: \mathbb{R}^m\mapsto \mathbb{R}^n
\end{equation}
Now we can define the discrete Laplace-Beltrami Operator conveniently as a mapping from the vertex space to the vertex space of the graph by consecutive application of the gradient and divergence operators:
\begin{equation}
	\underline{\Delta}=\overline{\nabla}^T \overline{\nabla}=\mathbf{L}:=\mathbf{D}\mathbf{G}=\mathbf{M}^T \mathbf{\Xi}^{-2}\mathbf{M}\in \mathbb{R}^{n\times n}, \underline{\Delta}=\overline{\nabla}^T \overline{\nabla}: \mathbb{R}^n\mapsto \mathbb{R}^n
\end{equation}
The inclusion of edge length reciprocals in this formulation makes it uniquely accurate in contrast to the commonly used Combinatorial Laplacian that is devoid of the metric dimension of the space modelled by the graph in question. 

The last differential operator to be introduced here is the Curl operator, that can measure rotations in a vector field. 
Here we assume, according to our proposed method, that vector fields are represented in the edge space of graphs; and naturally expect the curl vector field to be attributed to the face space of the graph. 
Considering this asuumption, it is easy to see that the curl operator requires an integral area element; and so we shall introduce the curl operator after our integral operators. 

By using the edge space or the hyper-edge space of voxel graphs or voxel complexs we can propose Discrete Spatial Integral Operators.
The first one is the Discrete Line/Curve-Integral Operator as a functional or a mapping from the Vertex space of the graph to the space of real numbers, which again uses the intermediate edge space of the graph for making the Riemann integrands:
\begin{equation}
	\underline{\int}=\boldsymbol{\mathfrak{s}}^{(1)}:=\frac{1}{2}\boldsymbol{\xi}^{T}\overline{\mathbf{M}}_{|E|,|V|}\in \mathbb{R}^{1\times n}, \underline{\int}=\boldsymbol{\mathfrak{s}}^{(1)}: \mathbb{R}^n\mapsto \mathbb{R}
\end{equation}
The second one is the Discrete Surface-Integral Operator defined as a mapping from the Vertex space of the graph to the space of real numbers through the intermediate Face space of the topological cell complex:
\begin{equation}
	\underline{\iint}=\boldsymbol{\mathfrak{s}}^{(2)}:=\frac{1}{4}\boldsymbol{\alpha}^{T}\overline{\mathbf{M}}_{|F|,|V|}\in \mathbb{R}^{1\times n}, \underline{\iint}=\boldsymbol{\mathfrak{s}}^{(2)}: \mathbb{R}^n\mapsto \mathbb{R}
\end{equation}
, in which $\boldsymbol{\alpha}=[\alpha_\varphi]_{|F|\times 1}$ is a vector containing the areas of the faces of the cell complex.
Note that for simplicial complexs the coefficient of equality would be $\frac{1}{3}$ in the above surface integral operator.
The third one is the Discrete Volume-Integral Operator defined as a mapping from the Vertex space of the graph to the space of real numbers through the intermediate Cell space of the topological cell complex:
\begin{equation}
	\underline{\iiint}=\boldsymbol{\mathfrak{s}}^{(3)}:=\frac{1}{8}\boldsymbol{\beta}^{T}\overline{\mathbf{M}}_{|C|,|V|}\in \mathbb{R}^{1\times n}, \underline{\iiint}=\boldsymbol{\mathfrak{s}}^{(3)}: \mathbb{R}^n\mapsto \mathbb{R}
\end{equation}
, in which $\boldsymbol{\beta}=[\alpha_\varphi]_{|C|\times 1}$ is a vector containing the volumes of the cells of the cell complex.

Similarly, the claim here is that these integral operators adequately approximate the continuous spatial integral operators, e.g.:
\begin{equation}
	\iiint_{\mathbf{x}\in \Omega} f(\mathbf{x})dV\approx \boldsymbol{\mathfrak{s}}^{(3)}\mathbf{f}
\end{equation}
, in which the $dV$ is the volume element of the continuous integral operator.

The curl operator requires an oriented mapping from the oriented edge space of the graph to its oriented face space. 
It is straight-forward to see and check for small graphs that the oriented Face-Edge Incidence Matrix of a hyper-graph (mesh) maps the Edge-Vertex Incidence vectors into a null space (zero vectors).
Denoting the Face-Edge Incidence matrix of a hyper graph as $\boldsymbol{\Omega}:=[\omega_{\varphi,\varepsilon}]_{|F|\times|E|}$, this means the following:
\begin{equation}
	\boldsymbol{\Omega}\mathbf{M}=\mathbf{0}_{|F|\times|V|}
\end{equation} 
In other words, $\boldsymbol{\Omega}^T$ is the null space (kernel) of the transposed Edge-Vertex Incidence Matrix of the same hyper-graph (after \cite{imperatore_topological_2016} and \cite{grady_discrete_2010}):
\begin{equation}
	\mathbf{M}^T\boldsymbol{\Omega}^T=\mathbf{0}_{|V|\times|F|}.
\end{equation} 

This practically means that the so-called cycle basis of the hypergraph (the elementary and irreducible faces of the mesh) that is represented by $\boldsymbol{\Omega}$ as  the solution to the following system of linear equations:
\begin{equation}
	\boldsymbol{\Omega}=\left(\mathbf{M}^T \backslash \mathbf{0}_{|V|\times|F|}\right)^T
\end{equation}
, where, $\mathbf{A}\backslash \mathbf{b}$ dentoes the solution to the linear equation $\mathbf{A}\mathbf{x}=\mathbf{b}$. 
Now, it is staright-forward to approximate the continuous curl operator on a hypergraph as a mapping from the edge space of the hypergraph to its face space obtained by applying $\boldsymbol{\Omega}$ as a linear map, i.e.:
\begin{equation}
	\boldsymbol{\nabla}\times \mathbf{F}\approx\boldsymbol{\EuScript{A}}^{-1}\boldsymbol{\Omega}\mathbf{F}
\end{equation} 
, where $\boldsymbol{\EuScript{A}}^{-1}$ is a diagonal matrix whose entries are the reciprocals of the face areas of the mesh, with the same indexing as the Face-Edge Incidence Matrix. 

Whilst the relations between such differential operators and the incidence matrices has been hinted to previously in some sources such as those referenced above, an exact algebraic treatment with the consideration of length, area, and volume normalizations seems to be missing in the litreature. 
It is easy to see that with these differential and integral operators Partial Differential Equations can be written elegenatly in an inherently discrete manner to be solved by linear algebraic solvers on complex (even non-manifold) spatial domains. 

In addition to the operators introduced thus far, the Jacobian operator (or the multivariate derivative) can be obtained by applying the gradient operator to a vector field attributed to the vertices of a hypergraph, i.e.:
\begin{equation}
	\boldsymbol{\EuScript{J}}\mathbf{F}:=\left[\mathbf{J}_{\iota}(\mathbf{f}_{\iota})\bigg|_{\mathbf{x}}\right]_{3n\times 3}:=\left[\left[\frac{\partial f_{\iota}}{\partial x_{i}}\right]_{3\times 3}\bigg|_{\mathbf{x}}\right]_{n\times 1}\approx \left(\mathbf{1}_{n\times 1}\otimes \mathbf{E}^T\right)\overline{\boldsymbol{\nabla}}\mathbf{F}
\end{equation}
, where $\mathbf{F}:=\left[\left(\mathbf{f}_{:,\varepsilon}\right)^T\right]_{m\times 3}$.

Note that the resultant Jacobian matrices will be listed as a block matrix or an array of $3\times 3$ matrices of size $3n\times 3$. 
This definition of the Jacobian matrices is useful when aiming to produce the best linear approximation of a vector function near the vertices of a mesh or hypergraph. 

The Hessian (the Jacobian of the gradient) of a scalar field can be obtained by firstly applying the gradient operator; producing the gradient vectors along the edges; applying the transposed edge-averaging operator (the onoriented Incidence matrix) to project back gradient vectors from edges to the vertices and then applying the gradient operator again, as follows:
\begin{equation}
	\boldsymbol{\EuScript{H}}\mathbf{f}:=\left[\mathbf{H}_{\iota}(f_{\iota})\bigg|_{\mathbf{x}}\right]_{3n\times 3}:=\left[\left[\frac{\partial^2 f_{\iota}}{\partial x_{i}\partial x_{j}}\right]_{3\times 3}\bigg|_{\mathbf{x}}\right]_{n\times 1}\approx \boldsymbol{\EuScript{J}}\mathbf{D}^{-1}\overline{\mathbf{M}}^T\boldsymbol{\EuScript{G}}\mathbf{E}
\end{equation}
, where $\boldsymbol{\EuScript{G}}:=diag(\overline{\boldsymbol{\nabla}}\mathbf{f})$ dennotes a diagonal matrix made up of the gradients attributed to the edges and $\mathbf{D}:=diag\left(diagonal\left(\mathbf{A}_{|V|\times |V|}\right)\right)$ denotes a diagonal matrix made up of the node degrees of the graph. 
The Hessian operator as such produces an array of $3\times 3$ matrices (a block matrix), each of whose elements is a Hessian matrix for the function value attributed to the corresponding vertex. 
Hessian matrices can be utilized in obtaining the Taylor approximations of scalar functions in Newton-type optimization procedures, which are better than the Jacobian approximations because they take the second derivative into account. 

Remarkably, our differential or integral operators do not even require the integration region to be a manifold.
In fact the line integral operator proposed here can even operate on networks of any genera.
Without loss of generality, for the higher-dimensional k-cells, in our derivation we particularly focus on the ``edge-space'' or the ``in-between space of vertices'', which we have so far introduced as interpolation space as the space to which we can conveniently attribute partial differentials and ``differential forms'' (integrands).
In the low-dimensional setting of the edge-space introduced here, we  derived these operators for the the physical dimension of the edges, faces, and cells between voxels that will be the Length, Area, and Volume of the dimensions $L$, $L^2$, and $L^3$, respectively, in terms of the elemental or base quantities in Physics (out of 7 elemental quantities, namely Mass, Length, Time, Electric Current, Absolute Temperature, Amount of Substance, and Luminous Intensity; see a brief introduction to Dimensional Analysis by \cite{Nourian_evaluation_2016}).

\section{Implementation: topoGenesis}

The workflow that has been put forth in this paper is implemented, for the large part, as a library of vectorized functions in the python programming language, released as the open-source package \href{https://genesis-lab.dev/products/topogenesis/}{topoGenesis} for computational geometry and topology operations in spatial computing and generative design (\href{https://github.com/shervinazadi/topoGenesis}{source available}) \cite{azadi2020topogenesis}.
The package topoGenesis is created on top of ubiquitous numeriacal and scientific python libraries such as Numpy \cite{harris2020array} and SciPy \cite{2020SciPy_NMeth} to ensure that the computational procedures are as accessible and efficient as possible.
Furthermore, python notebooks implementing the presented workflow with test cases are publicly accessible in \href{https://topogenesis.readthedocs.io/}{topoGenesis example workflows}.

\section{Conclusion: Application Outlook}
The shape of a spatial region affects the geodesic flows and the resultant geodesic distances within the manifold and thus it effectively influences almost any dynamic phenomenon or emergent pattern in space that is of practical interest in scientific and engineering applications. 
The configuration of a non-trivially shaped spatial region needs to be modelled as a discrete manifold for digital computing. 
In scientific and engineering applications (particularly in Computational Science and Engineering) the shape of the spatial region is almost always non-trivial, and so, one needs to go beyond using constructs defined in the context of continuous mathematics for spatial interpolation, differentiation, and integration. 

In summary, we can outline the two major application areas of the proposed methods as geo-spatial topological data analysis and geo-spatial simulations based on the so-called first principles encapsulated in PDE, Agent-Based Models or Cellular Automata.
For the first category of applications, the notion of manifold distance (geodesic distance from within the manifold) or network distance is key to defining metrics of similarity.
For the second category of applications, the notion of geodesic flow (of forces, electricity, fluids, pedestrians, and so on) is the central concept and often the goal of simulations to predict.

The proposed methods for construction of explicit topological models of spatial regions as graphs (technically as adjacency matrices or incidence matrices) pave the way for computing geodesics and geodesic distances on non-trivially shaped spatial regions.

Contrary to the common confusions and uncertainties typical to the topological analysis of data in the Euclidean space due to the ambigious choices about the notions of neighbourhood for constructing cell complexes, the proposed framework for utilizing cell complexes as their sparse graph representations not only effectively removes all such ambiguuities from the picture by putting forward only one explicit scale vector for defining the resolution of spatial analysis but also keeps all operations as efficient as operations on irregular simplicial complexes. 
Reflecting on this issue should ideally resolve the false dichotomy or the dilemma of image representations (typically assumed to be necessarily dense) and sparse representations associated with the simplicial complexes. 
In other words, the voxel-graph representations bring the best of the both worlds together coherently. 
Furthermore, the comprehensive spatial indexing scheme proposed here based on Morton Codes not only indexes the vertex space of the voxel graphs but also their edge spaces or even their higher-dimensional k-cell spaces elegantly and thus providing for unambiguous application of sparse matrices even in view of large-scale geo-spatial computations. 

Note that generalized Morton Codes defined here are globally unique identifiers of all subspaces of voxel complexes.
In this way the partial maps or models of space can be easily concatenated together without the need for special Euler operations for editing the graphs or meshes representing them. 
Algebraic computation of exact/deterministic metric geodesics and geodesic distances can be easily achieved by applying common graph traversal algorithms on the proposed graphs.
Alternatively, the stochastic or spectral counterparts of geodesics and geodesic distances known as random walks and diffusion distances can be computed very efficiently at scale on massive manifolds to enable simulations and spatial analyses (see the example of Google PageRank by \cite{page1999pagerank} and family of Random Walk models for such purposes in \cite{nourian2016configraphics}).
The proposed discrete differential and integral operators are unique in their algebraic simplicty and versatile efficacy and efficiency for spatial computing (computational spatial analysis and spatial simulation) in conjunction with the topological data models presented in Table\ref{tab:CombinatorialGraphs}. 

We can conclude by reflecting on a quote from the mathematician Doron Zeilberger:
"Conventional wisdom, fooled by our misleading "physical intuition", is that the real world is \emph{continuous}, and that discrete models are necessary evils for approximating the \emph{real} world, due to the innate discreteness of the digital computer."

\printbibliography
\pagebreak
\begin{appendices}
\section{Generalization} \label{app:Generalization}


Three of the more standardized neighbourhood descriptions within three-dimensional space are 6-neighbourhood (26-separating or conservative voxelization) (cf. \cite{zhangConservativeVoxelization2007}), 18-neighbourhood, and 26-neighbourhood (6-separating voxelization or thin voxelization).
Since 18-neighbourhood and 26-neighbourhood connectivity graphs can be constructed from 6-neighbourhood connectivity graph.
In this paper we present the mesh sampling algorithms (\ref{alg:MeshSurfaceSampling} \& \ref{alg:LineNetworkSampling}).
However, if the purpose of sampling is to construct a 26-neighbourhood, it is computationally cheaper to sample the mesh according to that connectivity type (see Table \ref{tab:ConnectivityType}).
\begin{table}[!htb]
	\footnotesize
	\caption{Relation of intersection objects with the most basic connectivity type (within 3D space) in the stencil}
	\centering
    \begin{tabularx}{\textwidth}{
    >{\raggedleft\arraybackslash\hsize=1.2\hsize}X
    >{\raggedright\arraybackslash\hsize=0.950\hsize}X
    >{\raggedright\arraybackslash\hsize=0.950\hsize}X
    >{\raggedright\arraybackslash\hsize=0.950\hsize}X
    >{\raggedright\arraybackslash\hsize=0.950\hsize}X
} 
    \toprule
    \textbf{Name} 
    & \textbf{Separation Type} 
    & \textbf{Connectivity Type} 
    & \textbf{Shared Element} 
    & \textbf{Intersection Origin}
    \\ 
    \toprule
    Conservative Voxelization 
    & 26-separation
    & 6-neighbourhood
    & Face
    & Corners of Voxel 
    \\ 
    \comment{
    - 
    & 18-separation 
    & 18-neighbourhood 
    & Edge
    & Middle of Voxel Edges 
    \\ 
    \hline}
    Thin Voxelization 
    & 6-separation 
    & 26-neighbourhood 
    & Vertex
    & Centroid of Voxel 
    \\ 
    \toprule
\end{tabularx}
	\label{tab:ConnectivityType}
\end{table}

\section{Notational Conventions} \label{app:NotationalConventions}
Our notational conventions are summarized in Table\ref{tab:NomenClature}.
\begin{table}[!htb]
	\footnotesize
	\caption{Notational Conventions}
	\centering
	\begin{tabularx}{\textwidth}
    {
        >{\small\raggedleft\arraybackslash\hsize=0.8\hsize}X
        >{\small\raggedright\arraybackslash\hsize=0.65\hsize}X
        >{\small\raggedright\arraybackslash\hsize=1.55\hsize}X
    } 
        \toprule
        \textbf{Object} 
        & \textbf{Notation} 
        & \textbf{Notes} 
        \\ 
        \toprule
        Scalars 
        & Lowercase Greek
        & Possibly Italic 
        \\ 
        Vectors 
        & Lowercase bold Latin
        & Possibly Italic 
        \\ 
        Matrices 
        & Uppercase bold Latin
        & Possibly Italic 
        \\ 
        Topological Sets
        & Uppercase Latins
        & as in $\EuScript{M}$$\rm{(V,E,F)}$
        \\ 
        Geometrical Sets
        & Uppercase Bold Fraktur
        & such as a list of voxels $\pmb{\mathfrak{V}}$
        \\ 
        Geometric Objects
        & Uppercase Euler Script
        & e.g. Plane $\EuScript{P}$, Mesh $\EuScript{M}$, and Bounding-Box $\EuScript{B}$.
        \\ 
        Universal Sets
        & Blackboard Maths Script
        & e.g. $\mathbb{R}^3$ and $\mathbb{Z}^3$ 
        \\ 
        \hline
        List 
        & [ ]
        & as in Python
        \\ 
        Tuple
        & ( )  
        & n-tuples as Tuples in Python
        \\ 
        Dictionary \& Set
        & \{ \}
        & as Dictionaries in Python
        \\ 
        \hline
        Round to nearest integer
        & $\nint{}$
        & e.g. the round function: $f(x)=\nint{x}$
        \\ 
        
        Hadamard Multiplication
        & $\odot$
        & Element-wise Array Multiplication a.k.a. Schur product
        \\ 
        Hadamard Division
        & $\oslash$
        & Element-wise Array Division a.k.a. Schur division
        \\ 
        Kronecker Product
        & $\otimes$
        & Outer product of block matrices
        \\ 
        Matrix Multiplication
        & Omited
        & e.g. $\mathbf{A}\mathbf{x}=\mathbf{b}$. Note that dot product is also denoted as matrix multiplication, i.e. $\langle \mathbf{a},\mathbf{b}\rangle=\mathbf{a}^T\mathbf{b}$
        \\ 
        \hline
        Object Property/Attribute
        & dot notation
        & in C-style pseudocodes e.g. $\mathbf{p}.x=\mathbf{p}[0]$
        \\ 
        Type Arguments
        & $ \langle\rangle$ 
        & in C-style pseudocodes e.g. $list\langle int\rangle$
        \\ 
        Scopes in Algorithms 
        & Indentation \& vertical rule
        & as  in Python's flow-control scopes syntax
        \\ 
        Array Items \& Slices
        & NumPy Fancy Indexing 
        & $\mathbf{A}[2,1]$ is the element at 2nd row and 3rd column,\linebreak 
        $\mathbf{A}[:,1]$ is the 2nd column, and \linebreak
        $\mathbf{A}[r,:]$ is an slice of $\mathbf{A}$ at rows $r={1,2}$ 
        \\ 
        
        \toprule
    \end{tabularx}
	\label{tab:NomenClature}
\end{table}

\end{appendices}

\end{document}